\documentclass[hidelinks]{aa}

\usepackage{CJKutf8}            
\usepackage{enumitem}           
\usepackage{orcidlink}          
\usepackage{sidecap}            
\usepackage{xcolor}             

\newcommand{\plus}{\raisebox{.15\height}{\scalebox{.75}{+}}}

\begin{document}

\title{Substructures induced by dust drag in protoplanetary disks}

\newcommand{\info}{\fnmsep\thanks{Corresponding author; \texttt{bi@mpia.de}}}

\author{
    Jiaqing Bi \begin{CJK*}{UTF8}{gbsn}(毕嘉擎)\end{CJK*} 
        \inst{1}\info\orcidlink{0000-0002-0605-4961} \and
    Mario Flock
        \inst{1}\orcidlink{0000-0002-9298-3029}      \and
    Dominik Ostertag
        \inst{1,2}\orcidlink{0009-0001-7800-7662}    \and
    Neele Lüttkemöller
        \inst{3}\orcidlink{0009-0009-4109-2283}      \and
    Sebastian Wolf
        \inst{3}\orcidlink{0000-0001-7841-3452}
}

\authorrunning{Jiaqing Bi et al.}

\institute{
    \inst{1} 
        Max-Planck-Institut für Astronomie, 
        69117 Heidelberg, Germany \\
    \inst{2}
        Fakultät für Physik und Astronomie, 
        Ruprecht-Karls-Universität Heidelberg, 
        69120 Heidelberg, Germany \\
    \inst{3} 
        Institut für Theoretische Physik und Astrophysik, 
        Christian-Albrechts-Universität zu Kiel, 
        24118 Kiel, Germany 
}

\date{Manuscript compiled on \today}

\abstract{
    Dust substructures observed in protoplanetary disks are commonly attributed to embedded planets; however, intrinsic gas--dust interactions can also generate complex morphologies. We performed two-dimensional, axisymmetric simulations of gas and dust that include dust back-reaction and parameterized turbulence to investigate how the streaming instability (SI) and vertical shear instability (VSI) shape dust distributions. With moderate viscosity and sufficiently high metallicity, we identify a characteristic shuttlecock-shaped dust substructure composed of a dense, vertically settled ``head'' and a vertically extended ``tail.'' This morphology arises from nonlinear SI driven by marginally coupled grains and the associated modification of gas flows. The dust scale height in the tail exceeds predictions based on the the simple diffusion--settling balance, indicating strong self-generated turbulence. With lower viscosity, VSI becomes more vigorous, disrupts midplane structures, and increases vertical stirring; nevertheless, for dust grains with Stokes numbers around 0.01, SI can still attain dust-to-gas ratios of up to 20--50, potentially approaching the Hill density for gravitational binding. Our results demonstrate that intrinsic gas--dust interactions can generate prominent dust substructures even in disks with finite viscosity and, under favorable conditions, concentrate dust to levels relevant for planetesimal formation.
}

\maketitle\nolinenumbers


\section{Introduction} \label{sec:intro}

Dust in protoplanetary disks plays a central role in planet formation and serves as one of the most direct observational tracers of disk dynamics. Observations with the Atacama Large Millimeter/submillimeter Array (ALMA) have revealed that dust emission commonly exhibits substructures, including annular gaps and rings (\citealt{almapartnership.CL.2015.07,andrews.SM.2016.04,andrews.SM.2018.12,isella.A.2016.12,long.F.2018.12,long.F.2019.09,cieza.LA.2019.01,perez.S.2019.07,bosschaart.Q.2026.04a}), central cavities (\citealt{vandermarel.N.2018.02,vandermarel.N.2021.01,vandermarel.N.2022.04,perez.S.2020.01,francis.L.2020.04,benisty.M.2021.07,facchini.S.2026.02}), winding spiral arms (\citealt{perez.LM.2016.09,dong.R.2018.06,huang.J.2018.12a,kurtovic.NT.2018.12,rosotti.GP.2020.06,yoshida.TC.2025.09}), and non-axisymmetric crescents and clumps (\citealt{fukagawa.M.2013.12,vandermarel.N.2013.06,isella.A.2018.12,casassus.S.2019.03,long.F.2022.09,ribas.A.2024.08,wolfer.L.2025.04}). The ubiquity of these features suggests that disks are intrinsically structured and dynamically active on planet-forming scales, potentially reflecting instabilities from coupled dust and gas. Because ALMA observations predominantly trace millimeter-sized grains, interpreting these substructures requires a detailed understanding of how marginally coupled dust is transported and concentrated within disks.

The dynamics of millimeter-sized grains is particularly important near the disk midplane, where dust back-reaction onto the gas is enhanced by vertical settling and where planetesimal formation is expected to initiate. However, grain growth beyond millimeter sizes faces several well-known barriers: rapid inward drift removes large grains on short timescales, high-velocity collisions lead to fragmentation, and inefficient sticking causes particles to bounce off rather than to grow (\citealt{weidenschilling.SJ.1977.07,blum.J.2008.09,brauer.F.2008.03,birnstiel.T.2010.04,guttler.C.2010.04,zsom.A.2010.04,zsom.A.2011.10,drazkowska.J.2023.07,birnstiel.T.2024.09}). Overcoming these barriers requires mechanisms that locally retain and concentrate dust, for example through trapping at pressure maxima or via instability-driven accumulation processes (\citealt{pinilla.P.2012.02,lyra.W.2013.09,dullemond.CP.2018.12,lesur.G.2023.07}). The observed substructures can therefore serve as key signposts of regions where dust accumulates efficiently enough to trigger planetesimal formation.

A leading mechanism for concentrating marginally coupled dust into dense structures is the streaming instability (SI; \citealt{youdin.AN.2005.02,johansen.A.2007.06}). The SI belongs to the broader class of resonant drag instabilities, which arise when dust streams through gas and the dust--gas drift resonates with a natural wave mode of the gas (\citealt{squire.J.2018.07,squire.J.2018.03}). Simulations demonstrate that SI can produce strong dust clumping and associated turbulence for millimeter-sized grains, and that its efficiency depends sensitively on the local dust-to-gas ratio and stopping time (\citealt{bai.XN.2010.10,carrera.D.2015.07,yang.CC.2017.10,lim.J.2024.07,lim.J.2025.03,ostertag.D.2025.03}). These properties make SI a natural bridge between disk dynamics and the onset of planetesimal formation.

Despite its effectiveness under favorable conditions, the growth and saturation of SI are sensitive to the gas dynamics. Turbulent diffusion and viscous damping can weaken or suppress SI by dispersing dust overdensities and disrupting the resonance that drives its growth (\citealt{chen.K.2020.03,umurhan.OM.2020.05,lim.J.2024.07}). This sensitivity highlights the need to consider SI in conjunction with other hydrodynamic processes that contribute to disk turbulence and angular momentum transport.

In regions with weak magnetic activity, hydrodynamic instabilities can provide an important source of turbulence. One such instability is the vertical shear instability (VSI), a baroclinic instability associated with the radial temperature gradient in disks with a short thermal relaxation time (\citealt{nelson.RP.2013.11,barker.AJ.2015.06,lin.MK.2015.09}). Simulations show that VSI generates turbulence and large-scale vertical gas motions that originate in the surface layers and extend toward the midplane, thereby modifying dust settling, concentration, and the conditions under which SI develops (\citealt{stoll.MHR.2014.12,stoll.MHR.2016.10,lin.MK.2017.11,lin.MK.2019.06}).

Recent studies have begun to explore the coexistence of SI and VSI in global disk models, revealing that dust back-reaction and hydrodynamic turbulence can mutually influence instability growth, saturation, and dust-layer structure. Using two-dimensional axisymmetric disk models with dust treated as Lagrangian particles, \citet{schafer.U.2020.03} showed that the outcome of SI--VSI interaction depends sensitively on their relative growth and saturation timescales: when VSI develops first, VSI-driven turbulence dominates the dust layer and inhibits settling, whereas when SI grows concurrently, dust back-reaction enhances midplane turbulence and produces denser dust concentrations. Building on this work, \citet{schafer.U.2022.10} showed that VSI-induced pressure perturbations can act as seeds for SI, enabling strong dust clumping at lower dust-to-gas ratios and smaller grain sizes than would be required for SI operating in isolation. More recently, \citet{schafer.U.2025.02} quantified dust diffusion driven by SI and VSI both separately and jointly, showing that VSI produces highly anisotropic, scale-dependent diffusion associated with large-scale motions, while SI contributes an intrinsic diffusion level that broadens the midplane dust layer. In parallel, \citet{huang.P.2025.06, huang.P.2025.06a} investigated SI--VSI coupling using both two- and three-dimensional multi-fluid simulations, highlighting the emergence of zonal flows, vortices, and secondary instabilities that further regulate dust concentration. In particular, \citet{huang.P.2025.06a} report that the outcome of SI--VSI interaction is independent of whether SI or VSI grows first, contrary to \citet{schafer.U.2020.03}. Collectively, these studies demonstrate that SI and VSI can mutually influence each other's saturation, turbulence properties, and dust morphology, motivating further exploration of their coupled behavior under more general disk conditions.

Existing studies of SI--VSI coexistence have largely focused on inviscid conditions. However, in realistic disk environments, SI and VSI are unlikely to be the only sources of angular momentum transport: disk winds, nonideal magnetohydrodynamic effects, and other processes can also contribute to the background viscosity (\citealt{bai.XN.2013.05,bai.XN.2013.08,bai.XN.2016.02,bethune.W.2017.04,lesur.G.2023.07}). Observational constraints suggest that $\alpha$ is approximately between $10^{-4}$ and $10^{-3}$, encompassing the processes discussed above (\citealt{rosotti.GP.2023.06}). Such additional turbulence can modify the development of instabilities; in particular, VSI is sensitive to turbulent diffusion, which can weaken its growth or shift unstable modes to smaller scales (\citealt{barker.AJ.2015.06,lin.MK.2015.09}). Further investigation is therefore required to better understand how these instabilities operate under more realistic disk conditions.

Motivated by these considerations, we investigated the coupled evolution of SI and VSI in viscous protoplanetary disks using two-dimensional axisymmetric models. We focused on the dynamics and morphology of marginally coupled millimeter-sized dust near the midplane. Our study explores how viscosity modifies the coupled evolution of SI and VSI, how dust substructures emerge under these conditions, and what morphological signatures can arise from the combined action of drag-driven clumping and vertical-shear turbulence. By extending SI--VSI studies into viscous regimes, our work provides a step toward linking instability-driven dust dynamics with the diversity of substructures observed in protoplanetary disks.

The structure of this work is as follows. We first describe the numerical methods and simulation setup in Sect. \ref{sec:methods}, and present our results on dust dynamics and substructure formation in viscous environments in Sect. \ref{sec:results}. We discuss the implications of our findings for disk observations in Sect. \ref{sec:discussion}, and conclude with a summary of our main results in Sect. \ref{sec:conclusion}.

\section{Method} \label{sec:methods}

We considered a two-dimensional, axisymmetric protoplanetary disk model composed of gas and dust orbiting a central star of mass $M_\star$. We use $\{r,\theta\}$ to denote the spherical radius and polar angle, and $\{R,Z\}$ for the cylindrical radius and height. Both coordinate systems are centered on the star. The superscript ``ini'' denotes initial values. The subscript ``0'' denotes evaluations in the midplane at the reference radius $(R = R_0, Z = 0)$. 

\subsection{Model of gas}

We assumed an isothermal equation of state
\begin{equation} 
    P = \rho_{\rm g}c_{\rm s}^2, 
\end{equation}
where $\{P,\,\rho_{\rm g},\,c_{\rm s}\}$ are the pressure, volumetric density, and sound speed of the gas, respectively. We assumed a vertically isothermal, time-independent temperature profile
\begin{equation} 
    T_{\rm g}(R) \propto (R/R_0)^{-q}. 
\end{equation}
Here, $q = 3/7$ is adopted from \citet{chiang.EI.1997.11}, leading to a moderately flared disk with an aspect ratio of the pressure-supported gas disk given by 
\begin{equation} 
    h_{\rm g}(R) \equiv H_{\rm g}/R = h_{\rm g0}(R/R_0)^{2/7}.
\end{equation} In the above, we assume $h_{\rm g0} = 0.1$, 
\begin{equation} 
    H_{\rm g}(R) = c_{\rm s}/\Omega_{\rm K} 
\end{equation}
is the pressure scale height of the gas,
\begin{equation} 
    \Omega_{\rm K}(R) = v_{\rm K}/R = \sqrt{GM_\star / R^3} 
\end{equation}
is the Keplerian angular velocity, and $G$ is the gravitational constant. The disk's self-gravity and magnetic fields are neglected. 

\subsection{Model of dust}

We considered a single species of dust modeled as a pressureless fluid. The dynamical coupling between gas and dust is parameterized by the Stokes number
\begin{equation} 
    {\rm St} = \tau_{\rm s}\Omega_{\rm K}, 
\end{equation}
where $\tau_{\rm s}$ is the particle stopping time characterizing the frictional drag force between gas and dust. For dust grains that remain marginally coupled to the gas, we assumed the gas--dust interaction is in the Epstein regime. Assuming a fixed size ($s$) and internal density ($\rho_\circ$), the stopping time then becomes \citep{weidenschilling.SJ.1977.07}
\begin{equation} 
    \tau_{\rm s} = \rho_\circ s / \rho_{\rm g}c_{\rm s}, 
\end{equation} 
and in practice, it is prescribed by
\begin{equation} 
    \tau_{\rm s} = \frac{\rho_{\rm g0}^{\rm ini}}{\rho_{\rm g}}\frac{c_{\rm s0}}{c_{\rm s}}\frac{{\rm St}_{\rm 0}^{\rm ini}}{\Omega_{\rm K0}}. 
\end{equation}
We adopted a fiducial reference Stokes number ${\rm St}_0^{\rm ini} = 10^{-3}$, which corresponds to approximately 100-$\mu$m-sized grains at $R = 100 \,{\rm au}$ in protoplanetary disks such as HL Tau\footnote{Since the global $\rho_{\rm g}$ profile does not significantly deviate from its initial condition and $c_{\rm s}$ is time-independent, we hereafter use ${\rm St}_0$ to refer to ${\rm St}_0^{\rm ini}$ for brevity.}\textsuperscript{,}\footnote{Assuming a disk mass of 0.2 $M_\odot$ and an outer disk radius of 150 au \citep{booth.AS.2020.03}.}. We also explore the effect of varying the Stokes number in Sect. \ref{sec:compare_grain_size}.

\subsection{Basic equations}

The hydrodynamic equations for gas and dust are given by 
\begin{align} 
    &\frac{\partial \rho_{\rm g}}{\partial t} + \nabla \cdot (\rho_{\rm g} \textit{\textbf{V}}_{\rm g}) = 0 \\[0.5em]
    &\begin{aligned}
        \frac{\partial \textit{\textbf{V}}_{\rm g}}{\partial t} + \textit{\textbf{V}}_{\rm g} \cdot \nabla \textit{\textbf{V}}_{\rm g} = - &\nabla \Phi + \frac{\epsilon}{\tau_{\rm s}}(\textit{\textbf{V}}_{\rm d} - \textit{\textbf{V}}_{\rm g}) \\[0.5em]
        - &\frac{1}{\rho_{\rm g}} \nabla P + \frac{1}{\rho_{\rm g}} \nabla \cdot \mathcal{T}
    \end{aligned} \\[0.5em]
    &\frac{\partial \rho_{\rm d}}{\partial t} + \nabla \cdot (\rho_{\rm d} \textit{\textbf{V}}_{\rm d}) = 0 \\[0.5em]
    &\frac{\partial \textit{\textbf{V}}_{\rm d}}{\partial t} + \textit{\textbf{V}}_{\rm d} \cdot \nabla \textit{\textbf{V}}_{\rm d} = - \nabla \Phi - \frac{1}{\tau_{\rm s}}(\textit{\textbf{V}}_{\rm d} - \textit{\textbf{V}}_{\rm g}).
\end{align}
Here, $\{\rho_{\rm g},\textit{\textbf{V}}_{\rm g}\}$ and $\{\rho_{\rm d},\textit{\textbf{V}}_{\rm d}\}$ are the volumetric density and velocity vector of gas and dust, respectively. 
\begin{equation} 
    \Phi(r) = -GM_\star/r 
\end{equation}
is the gravitational potential from the star, and
\begin{equation} 
    \epsilon \equiv \rho_{\rm d}/\rho_{\rm g} 
\end{equation} 
is the dust-to-gas density ratio. The viscous stress tensor is given by
\begin{equation} 
    \mathcal{T} = \rho_{\rm g}\nu\left[\nabla\textit{\textbf{V}}_{\rm g} + (\nabla\textit{\textbf{V}}_{\rm g})^{\dagger} - \frac{2}{3}\textit{\textbf{I}}\,\nabla \cdot\textit{\textbf{V}}_{\rm g}\,\right],
\end{equation}
where $\textit{\textbf{I}}$ is the identity tensor, and the kinematic viscosity $\nu$ is parameterized as 
\begin{equation} 
    \nu = \alpha c_{\rm s} H_{\rm g}.
\end{equation}
In our models, $\alpha$ is a dimensionless parameter for viscosity (\citealt{shakura.NI.1973.01}) that is spatially constant unless otherwise specified (e.g., in Sect. \ref{sec:compare_alpha_vert}). The value of $\alpha$ varies between models, and we primarily focused on disk models with $\alpha$ between $10^{-5}$ and $10^{-4}$. But we also explore the effect of varying $\alpha$ in Sect. \ref{sec:compare_alpha_glob}. Dust back-reaction onto the gas is included in our models, while dust diffusion in the turbulent gas is intentionally neglected in order to isolate the dust stirring effect driven by dust back-reaction\footnote{A second motivation is that the dust diffusion prescription in the \textsc{fargo3d} code cannot account for dust back-reaction and therefore does not strictly conserve the angular momentum of the gas \citep{weber.P.2018.02}, complicating a consistent treatment of dust feedback in turbulent flows.}.

\subsection{Gas and dust initialization}

The volumetric density of the gas ($\rho_{\rm g}$) is initialized using the vertically integrated surface density profile
\begin{equation} \label{eq:sigmag}
    \Sigma_{\rm g}^{\rm ini}(R) = \Sigma_{\rm g0}^{\rm ini}\left[(R/R_0)+(R/R_{\rm c})^5\right]^{-1},
\end{equation}
where $\Sigma_{\rm g0}^{\rm ini}$ can be arbitrary for our non-self-gravitating disk models, and the corresponding volumetric density is
\begin{equation} 
    \rho_{\rm g}^{\rm ini} = \frac{\Sigma_{\rm g}^{\rm ini}}{\sqrt{2\pi}H_{\rm g}} \times \exp{\left(-\frac{Z^2}{2H_{\rm g}^2}\right)}. 
\end{equation}
The last term of Eq.~\ref{eq:sigmag} provides a steeper decrease in the gas density profile beyond a cut-off radius $R_{\rm c}$, which is inspired by recent rotation curve studies of protoplanetary disks (e.g., \citealt{martire.P.2024.06}). With $R_{\rm c} = 2R_0$, the deviation between Eq.~\ref{eq:sigmag} and an ordinary power-law profile $\Sigma_{\rm g}\propto R^{-1}$ becomes noticeable (about a factor of 2) starting from $2R_0$, and reaches a factor of roughly 20 at $5R_0$.

The volumetric density of the dust ($\rho_{\rm d}$) is initialized via the dust-to-gas density ratio
\begin{equation} 
    \epsilon^{\rm ini} \equiv \rho_{\rm d}^{\rm ini}/\rho_{\rm g}^{\rm ini} = \epsilon_0^{\rm ini} \times \delta_R \times \delta_Z.
\end{equation}
Here, $\delta_R$ is a radial-direction taper given by
\begin{equation}
    \delta_R = 
    \begin{cases}
        \left(\dfrac{1}{1-\zeta}\dfrac{R - R_{\rm min}}{R_0 - R_{\rm min}}\right)^2 \\[1.5em]
        \qquad\qquad\qquad {\rm for}\; R < [R_0-\zeta(R_0 - R_{\rm min})], \\[0.5em]
        \left(\dfrac{1}{1-\zeta}\dfrac{R - R_{\rm max}}{R_0 - R_{\rm max}}\right)^2 \\[1.5em]
        \qquad\qquad\qquad {\rm for}\; R > [R_0-\zeta(R_0 - R_{\rm max})], \\[0.5em]
        1.0 \qquad\qquad\;\:\:\text{for anywhere else},
    \end{cases} 
\end{equation}
where $R_{\rm min}$ and $R_{\rm max}$ are the inner and outer radial boundaries of the computational domain (specified in Sect. \ref{sec:numerics}), and $\zeta = 3/4$, such that $\rho_{\rm d}^{\rm ini}$ gradually becomes zero when approaching the radial boundaries. Meanwhile, $\delta_Z$ is a vertical-direction taper given by
\begin{equation} 
    \delta_Z = \exp{\left(-Z^2\frac{H_{\rm g}^2 - H_{\rm d}^2}{2H_{\rm g}^2 H_{\rm d}^2}\right)},
\end{equation} 
such that $\rho_{\rm d}^{\rm ini}$ retains a vertically Gaussian profile with a dust scale height $H_{\rm d}^{\rm ini} = 0.1H_{\rm g}^{\rm ini}$. We adopted this pre-settled dust layer to reduce computational cost by bypassing the initial dust-settling phase, which is not expected to affect the subsequent dust evolution in our models. With a fiducial value $\epsilon_0^{\rm ini} = 1$, this setup corresponds to a disk model with a total dust-to-gas mass ratio (i.e., metallicity) of $\mathcal{Z} \approx 0.1$. We also explore the effect of varying $\mathcal{Z}$ in Sect. \ref{sec:compare_mass_ratio}.

The azimuthal velocities are initialized to 
\begin{align}
    & V_{\rm g,\phi}^{\rm ini} = R\Omega_{\rm K}(R)\sqrt{1-2\eta}, \\[0.5em]
    & V_{\rm d,\phi}^{\rm ini} = r\Omega_{\rm K}(r),
\end{align}
where
\begin{equation} 
    \eta = \frac{1}{2}\left[(p+q)h_{\rm g}^2 + q\left(1 - \frac{R}{r}\right)\right]
\end{equation}
is a term from the radial gradient of gas pressure, with
\begin{align}
\begin{split}
    p & = -\frac{{\rm d\,log}\,\rho_{\rm g}|_{Z=0}}{{\rm d\,log}\,R} = -\frac{{\rm d\,log}\,\Sigma_{\rm g}}{{\rm d\,log}\,R} + \frac{{\rm d\,log\,{\it H}_{g}}}{{\rm d\,log}\,R} \\[0.5em]
    & = \frac{(R/R_0)+5(R/R_{\rm c})^5}{(R/R_0)+\phantom{5}(R/R_{\rm c})^5} + \frac{9}{7} \\[0.5em]
    & = \frac{5(R/R_0)^4 + 2^5}{\phantom{5}(R/R_0)^4 + 2^5} + \frac{9}{7}.
\end{split}
\end{align}
Both radial and vertical velocities of gas and dust are initialized to zero. We note that this initialization does not correspond to a strict steady state of the coupled gas--dust system (\citealt{kanagawa.KD.2017.08}). In the absence of explicit dust diffusion, the vertical equilibrium of the dust layer cannot be maintained, as dust continues to settle and therefore immediately deviates from the initial state once the simulation begins. Consequently, even an exact steady-state initialization would not be preserved. We therefore adopted this setup as a convenient starting configuration and do not expect the velocity initialization to significantly affect the subsequent dust evolution.

\subsection{Numerical tool} \label{sec:numerics}

Our disk models are evolved by the multi-fluid hydrodynamic code \textsc{fargo3d} (\citealt{benitez-llambay.P.2016.03, benitez-llambay.P.2019.04}). We adopted an axisymmetric domain in spherical coordinates, spanning $0.1R_0$ to $5.0R_0$ in $r$, and between $\pi/2 \pm {\rm atan}(2h_{\rm g0})$ in $\theta$ (i.e., roughly $\pm 2H_{\rm g}$ in the vertical direction). Our fiducial grid resolution is $N_r \times N_\theta = 8{,}000\times 800$ with logarithmic spacing in $r$ and uniform spacing in $\theta$. This corresponds to approximately 200 cells per $H_{\rm g}$ in both directions. We also explore the effect of varying the grid resolution in Appendix \ref{sec:compare_resolution}.

We adopted a wave-killing (i.e., Stockholm) boundary condition for gas in the radial direction. The inner (respectively, outer) damping zone ends (begins) at the radius where the Keplerian velocity is 1.5 times slower (faster) than that of the inner (outer) radial boundary. At the boundaries, the gas density is assumed to be in hydrostatic equilibrium, and the dust density is assumed to be symmetric. The meridional velocities of both gas and dust are assumed to be zero at the boundaries, except that the dust is allowed to leave the computational domain through the inner radial boundary.

\section{Result} \label{sec:results}

In this section we first describe the dust dynamics in two of our fiducial setups and then explore a broader parameter space to understand how dust dynamics and the resulting substructures depend on dust and gas properties.

\subsection[Dust distribution in the model with alpha = 1e-4]{Dust distribution in the model with $\alpha = 10^{-4}$} \label{sec:stream_lines_1e-4a}

\begin{figure*}
\sidecaption
\includegraphics[width = 12cm]{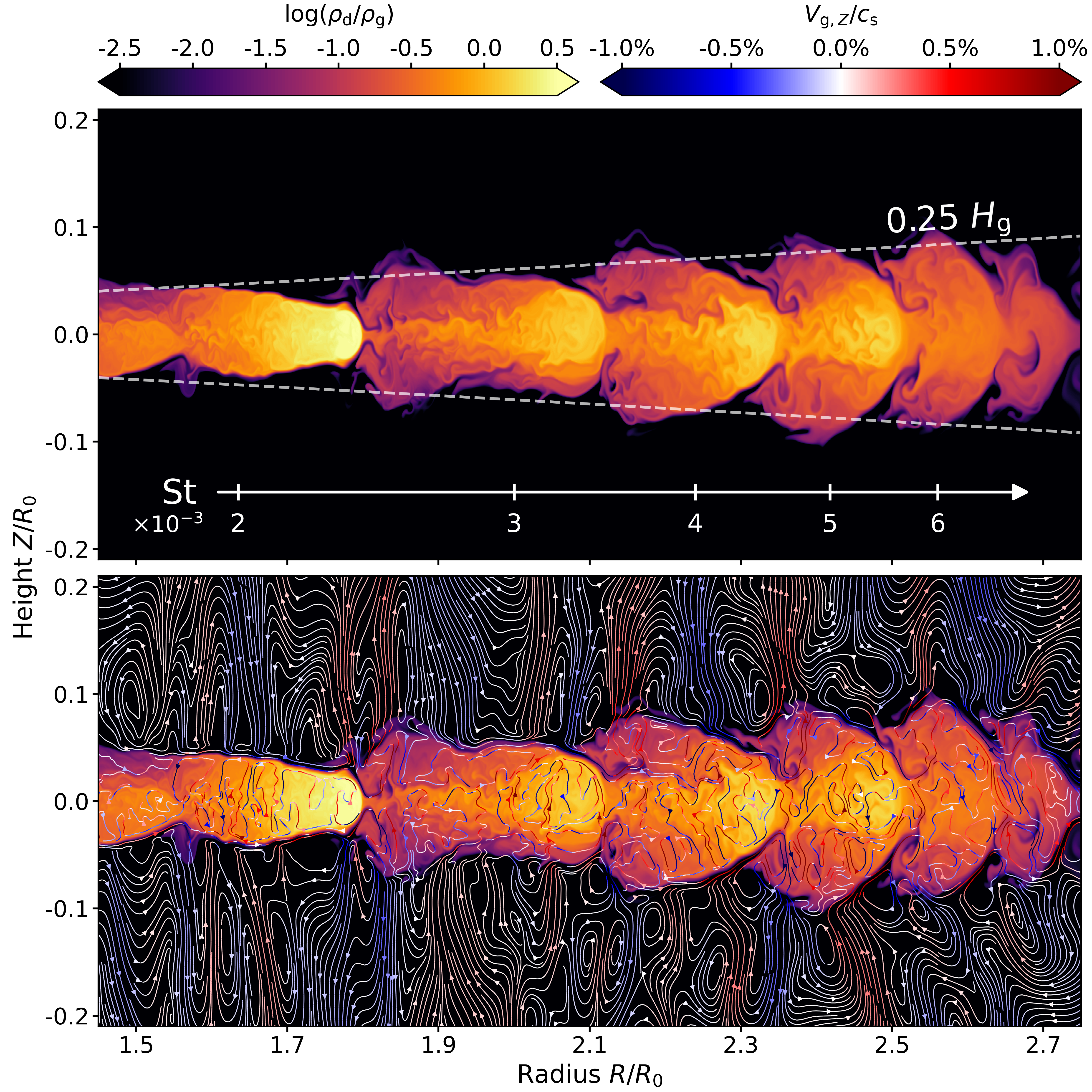}
\caption{Dust-to-gas density ratio at $t = 400\,T_0$ from the model with $\alpha = 10^{-4}$ and ${\rm St}_0 = 10^{-3}$. Dashed curves in the upper panel denote contours for 25\% of the gas scale height, and the arrow below denotes the Stokes number of dust grains in the midplane. Gas streamlines are overplotted in the lower panel, with the line color denoting the vertical gas velocity.
\label{fig:stream_lines_1e-4a}}
\end{figure*}

\begin{figure}[t]
\centering
\includegraphics[width = \columnwidth]{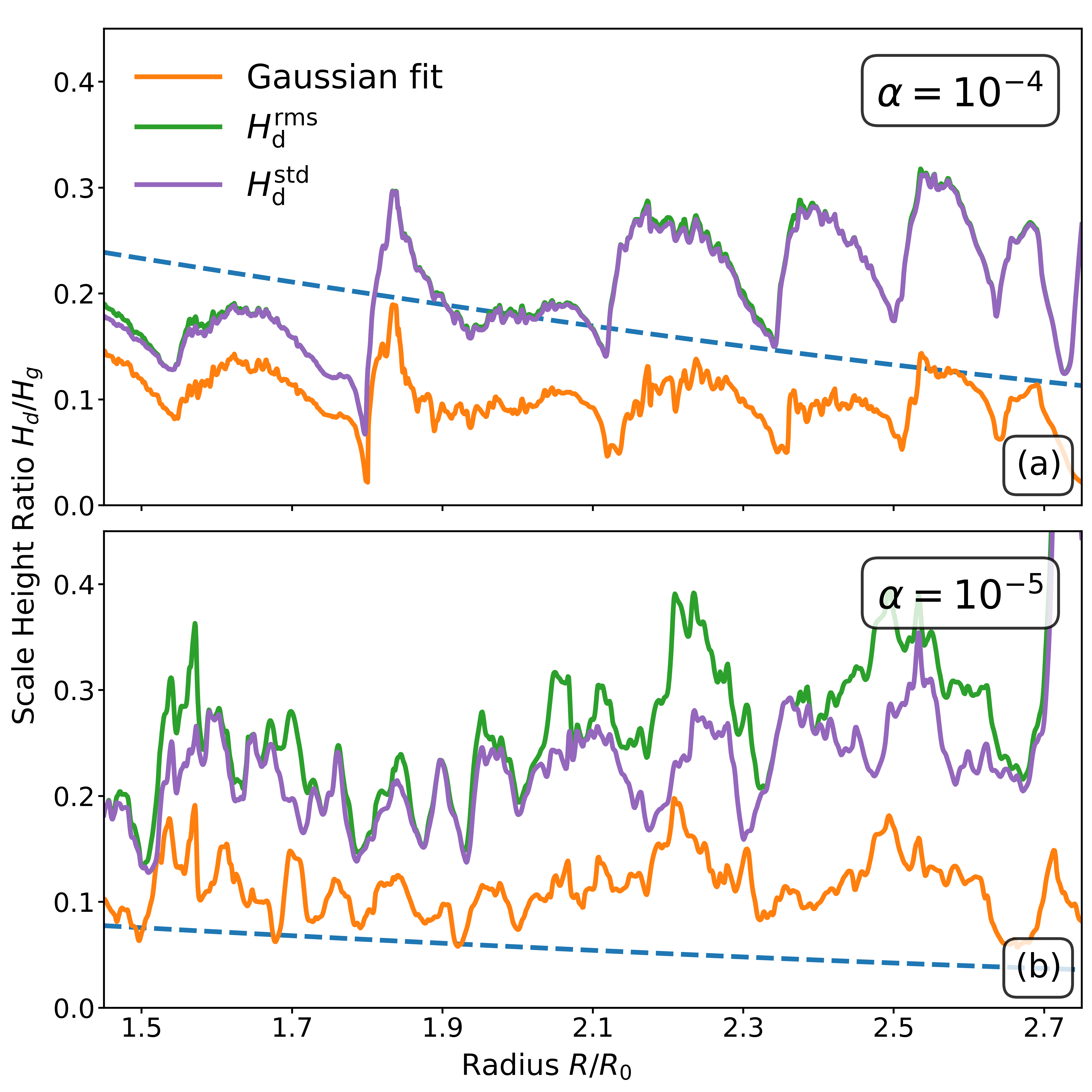}
\caption{Radial profiles of the dust-to-gas scale height ratio at $t = 400\,T_0$ from models with $\alpha = 10^{-4}$ (panel~$a$) and $\alpha = 10^{-5}$ (panel~$b$), both with ${\rm St}_0 = 10^{-3}$. The dust scale heights are measured using the root mean square ($H_{\rm d}^{\rm rms}$, Eq.~\ref{eq:hd_rms}), the standard deviation ($H_{\rm d}^{\rm std}$, Eq.~\ref{eq:hd_std}), or by fitting a Gaussian profile. The dashed curve indicates a reference-only expected ratio assuming explicit dust diffusion, as given by Eq.~\ref{eq:hd_alpha}. The radial extent is the same as in Figs.~\ref{fig:stream_lines_1e-4a} and \ref{fig:stream_lines_1e-5a}.
\label{fig:scale_height_1e-3s}}
\end{figure}

Figure~\ref{fig:stream_lines_1e-4a} shows the distribution of dust-to-gas density ratio at 400 reference orbital periods\footnote{The reference orbital period is given by $T_0 = 2\pi/\Omega_{\rm K0}$.} in the disk model with $\alpha = 10^{-4}$ and ${\rm St}_0 = 10^{-3}$. For the first time, we demonstrate that strong dust back-reaction from marginally coupled grains in the midplane can lead to the formation of spatially episodic, shuttlecock-shaped dust substructures that feature a denser, more vertically settled ``head,'' and a less dense but more vertically extended ``tail.'' 

The head of the shuttlecock exhibits a peak dust-to-gas density ratio of typically $1 \lesssim {\rm max}(\rho_{\rm d}/\rho_{\rm g}) \lesssim 3$, and therefore does not qualify as strong clumping. This is consistent with expectations, as our adopted metallicity $\mathcal{Z} = 0.1$ lies below the threshold required for strong clumping due to SI \citep{lim.J.2024.07}:
\begin{align}
\begin{split} \label{eq:z_crit}
\log \mathcal{Z}_{\rm crit}(\alpha, \tau_{\rm s}) 
= \: &0.15\times\log\alpha\times\log\alpha \,+ 1.18\times\log\alpha \\
- \: &0.24\times\log\alpha\times\log\tau_{\rm s} - 1.48\times\log\tau_{\rm s}.
\end{split}
\end{align}
For dust grains with ${\rm St} = 3\times10^{-3}$ at $R = 2R_0$, the critical value is $\mathcal{Z}_{\rm crit} \gtrsim 0.17$ in our setup\footnote{\citet{lim.J.2024.07} did not include explicit dust diffusion, but they accounted for dust self-gravity. Consequently, the threshold for strong clumping in our setup may be even higher.}.

To quantify the vertical extent of the dust layer, we used the mass-weighted root mean square vertical height
\begin{equation} \label{eq:hd_rms}
    H_{\rm d}^{\rm rms} = \sqrt{\frac{\sum(m_{\rm d}Z^2)}{\sum m_{\rm d}}}
\end{equation}
for the dust scale height relative to the midplane, and the weighted standard deviation
\begin{equation} \label{eq:hd_std}
    H_{\rm d}^{\rm std} = \sqrt{\frac{\sum(m_{\rm d}Z^2)}{\sum m_{\rm d}} - \left(\frac{\sum(m_{\rm d}Z)}{\sum m_{\rm d}}\right)^2}
\end{equation}
for the dust scale height relative to the center of the dust layer. The measurements are shown in Fig.~\ref{fig:scale_height_1e-3s}$a$. We find that Eqs.~\ref{eq:hd_rms} and \ref{eq:hd_std} yield similar results, meaning that the dust layer is close to being centered at the disk midplane. We further performed fits with a Gaussian centered at the midplane, but the result deviates from $H_{\rm d}^{\rm rms}$ and $H_{\rm d}^{\rm std}$, meaning that the dust layer does not match a Gaussian profile. Therefore, we adopted $H_{\rm d}^{\rm std}$ as our fiducial measurement of dust scale height hereafter. Without explicit dust diffusion, the dust scale height reaches $H_{\rm d}/H_{\rm g} \gtrsim 0.15$ at the head of the shuttlecocks and $H_{\rm d}/H_{\rm g} \gtrsim 0.25$ at the tail. Notably, these ratios are comparable to, or even above, the expected value resulting from the balance between vertical settling and turbulent diffusion (e.g., \citealt{dubrulle.B.1995.04, youdin.AN.2007.12}; see Fig.~\ref{fig:scale_height_1e-3s}):
\begin{equation} \label{eq:hd_alpha}
    \frac{H_{\rm d}}{H_{\rm g}} = \left(1 + \frac{\rm St}{\alpha}\frac{\rm 1+2St}{\rm 1+St}\right)^{-1/2}.
\end{equation}
However, we emphasize that this value should be regarded only as a reference estimate, since explicit dust diffusion is not included in our models and a true settling-diffusion equilibrium is therefore not expected to be established.

\subsubsection*{Why is the substructure shuttlecock-shaped?} \label{sec:compare_alpha_vert}

\begin{figure}[t]
\centering
\includegraphics[width = \columnwidth]{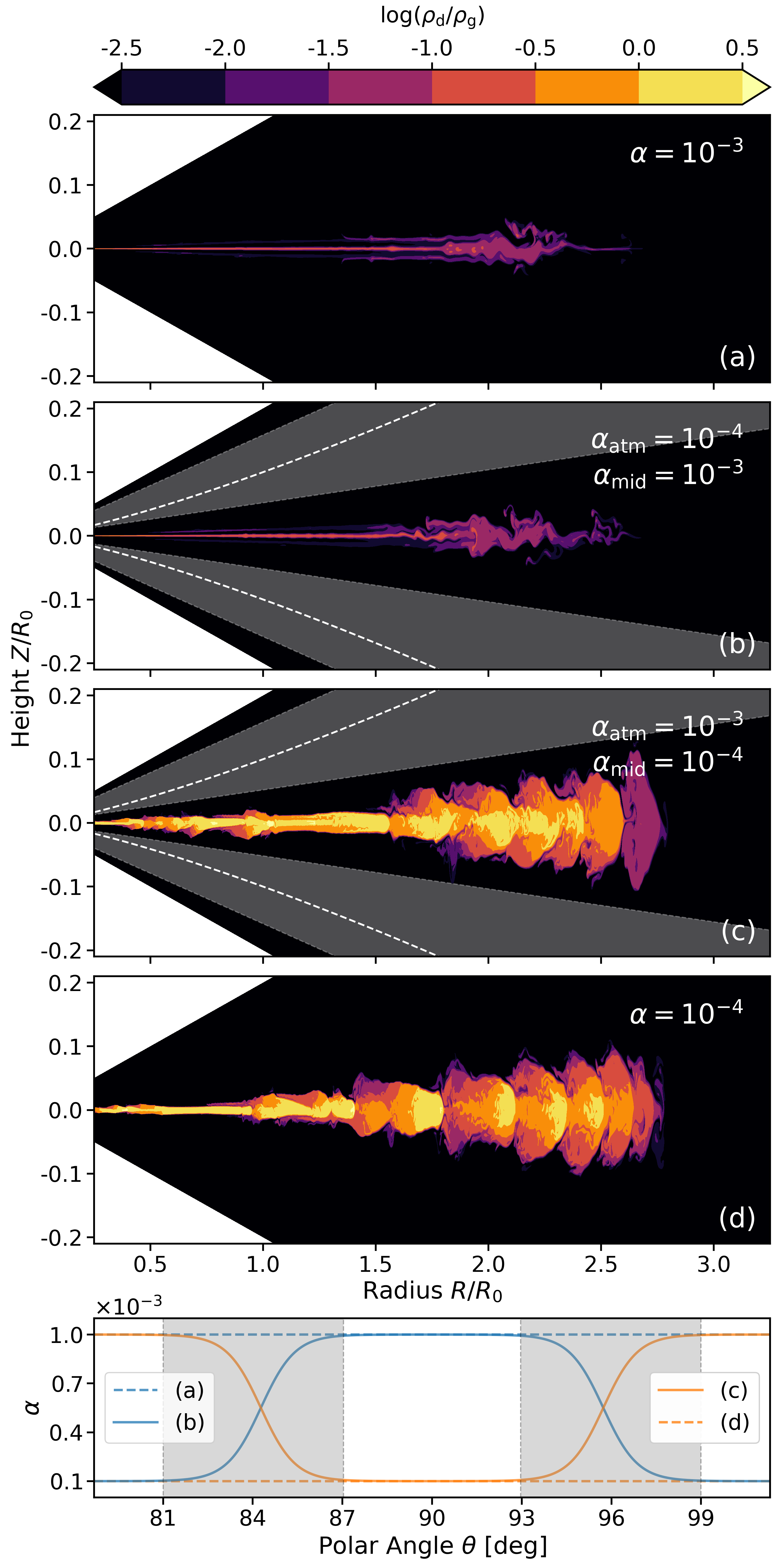}
\caption{Dust-to-gas density ratio at $t = 400\,T_0$ from models with different vertical prescriptions of the $\alpha$ parameter. Panel~$a$: Model in which $\alpha = 10^{-3}$. Panel~$b$: Model in which $\alpha$ decreases from $10^{-3}$ in the midplane to $10^{-4}$ in the atmosphere. Panel~$c$: Model in which $\alpha$ increases from $10^{-4}$ in the midplane to $10^{-3}$ in the atmosphere. Panel~$d$: Model in which $\alpha = 10^{-4}$. Last panel: Corresponding vertical profiles of $\alpha$, given by Eqs.~\ref{eq:window1}--\ref{eq:sigmoid2}. Dashed curves in panels~$b$ and $c$ denote contours for one gas scale height. The shaded regions indicate where the transition of $\alpha$ occurs and are bounded by aspect ratios evaluated at the inner and outer radial boundaries.
\label{fig:compare_alpha_vert}}
\end{figure}

The streamlines of gas velocity are overplotted in the bottom panel of Fig.~\ref{fig:stream_lines_1e-4a}. The streamlines exhibit radially alternating, vertically extended gas flows that appear to compress the head toward the midplane and stretch the tail in the opposite direction. However, these alternating gas flows may either cause the shuttlecock or be its consequence. On the one hand, the gas flows might be the breathing modes of VSI that are symmetric about the midplane, and thus the cause of the shuttlecock, provided that the midplane-crossing corrugation modes are suppressed by dust back-reaction (\citealt{lin.MK.2017.11, lin.MK.2019.06, schafer.U.2020.03, huang.P.2025.06}). On the other hand, the gas flows might be the result of SI in the midplane, as has also been observed in Fig.~9 of \citet{schafer.U.2020.03}, and thus the consequence. To distinguish between these two scenarios, we first performed one control simulation with a higher viscosity of $\alpha = 10^{-3}$, which was accompanied by two additional runs in which $\alpha$ was allowed to vary between $10^{-4}$ and $10^{-3}$ along the vertical direction. The results are shown in Fig.~\ref{fig:compare_alpha_vert}.

In the models where $\alpha$ varies spatially, the vertical profiles of $\alpha$ are given by
\begin{align} 
    \label{eq:window1}
    \alpha_1(\theta) &= \phantom{1}(9\psi_\theta-8) \times 10^{-4} \\
    \label{eq:window2}
    \alpha_2(\theta) &= (19-9\psi_\theta) \times 10^{-4}, 
\end{align}
where $\psi_\theta = \psi^+(\theta) + \psi^-(\theta) \in (1,2)$ is a window function connected by two mirrored logistic sigmoids
\begin{align}
    \label{eq:sigmoid1}
    \psi^+(\theta) = \phantom{1\,-} &\left[1 + \exp{\left(-\frac{\theta-\pi/2+10\delta_\theta}{\delta_\theta}\right)}\right]^{-1} \\
    \label{eq:sigmoid2}
    \psi^-(\theta) = 1\,- &\left[1 + \exp{\left(-\frac{\theta-\pi/2-10\delta_\theta}{\delta_\theta}\right)}\right]^{-1}, 
\end{align}
with $\delta_\theta = {\rm arctan}(0.1\times h_{\rm g0})$. This enables $\alpha$ to transition smoothly within a vertical range around $Z \sim H_{\rm g}$ (see the last panel in Fig.~\ref{fig:compare_alpha_vert}), thereby not introducing sharp gradients of $\alpha$ close to the midplane that would impact dust dynamics there. 

With a globally $\alpha = 10^{-3}$ prescription (Fig.~\ref{fig:compare_alpha_vert}$a$), the shuttlecock-shaped substructure disappears as a result of the suppression of both SI and VSI. The shuttlecock structure does not reappear when $\alpha$ decreases to $10^{-4}$ only in the disk atmosphere (panel~$b$), indicating that VSI, if present in the globally $\alpha = 10^{-4}$ model (panel~$d$), is unlikely to be responsible for the formation of the shuttlecock. In fact, as we show in Fig.~\ref{fig:compare_specificam} and Appendix~\ref{sec:compare_alpha_long}, VSI remains inactive when $\alpha \geq 10^{-4}$ in the disk atmosphere. In contrast, the shuttlecock re-emerges when $\alpha$ decreases to $10^{-4}$ in the midplane while remaining large in the atmosphere (panel~$c$). This behavior suggests that the shuttlecock is the result of SI, i.e., strong dust back-reaction from marginally coupled dust grains in the midplane. 

\subsection[Dust distribution in the model with alpha = 1e-5]{Dust distribution in the model with $\alpha = 10^{-5}$} \label{sec:stream_lines_1e-5a}

\begin{figure*}
\sidecaption
\includegraphics[width = 12cm]{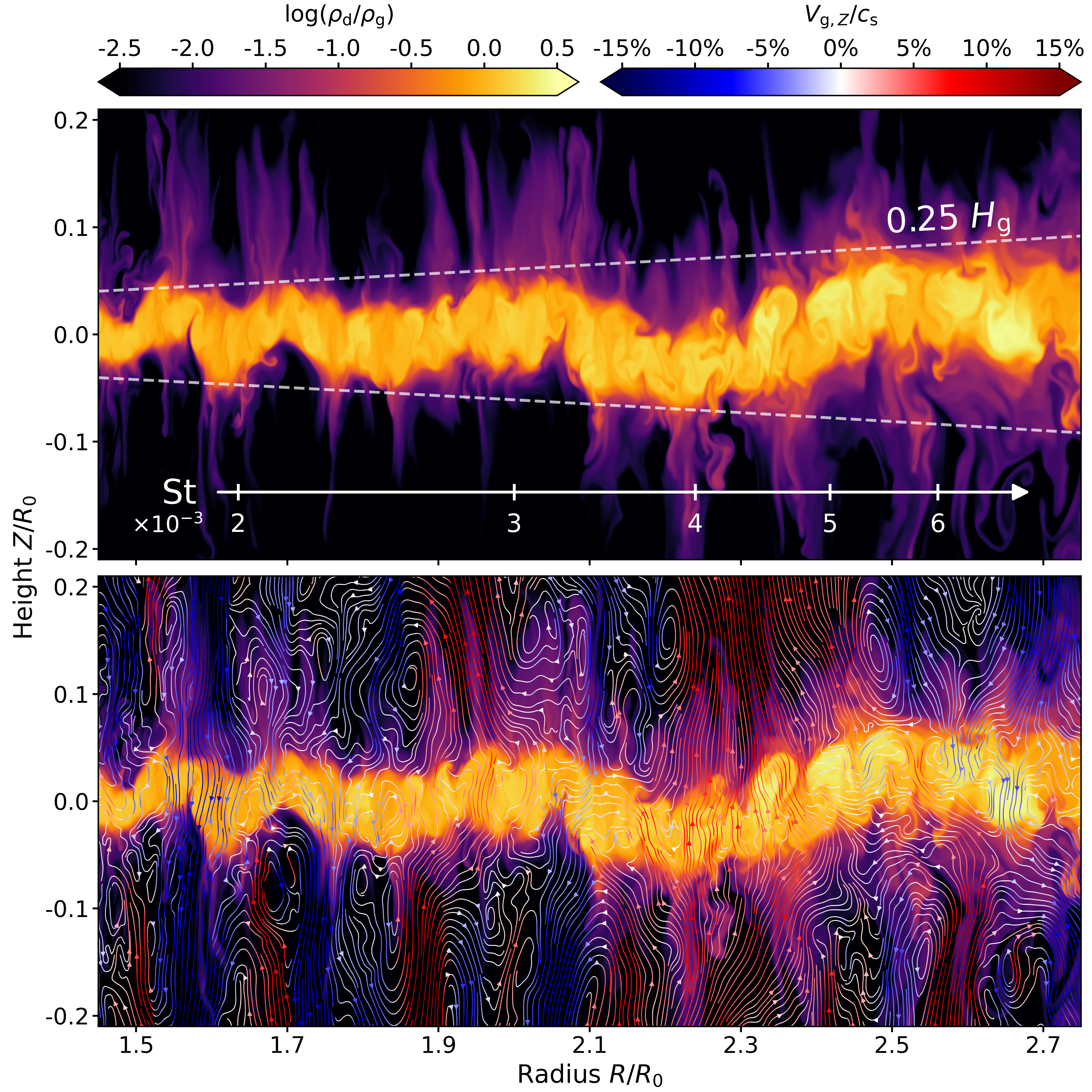}
\caption{Similar to Fig.~\ref{fig:stream_lines_1e-4a}, but from the model with $\alpha = 10^{-5}$. The colorbar for gas streamlines has been adjusted to fit stronger vertical gas flows in this lower-viscosity disk model.
\label{fig:stream_lines_1e-5a}}
\end{figure*}

\begin{figure}[t]
\centering
\includegraphics[width = \columnwidth]{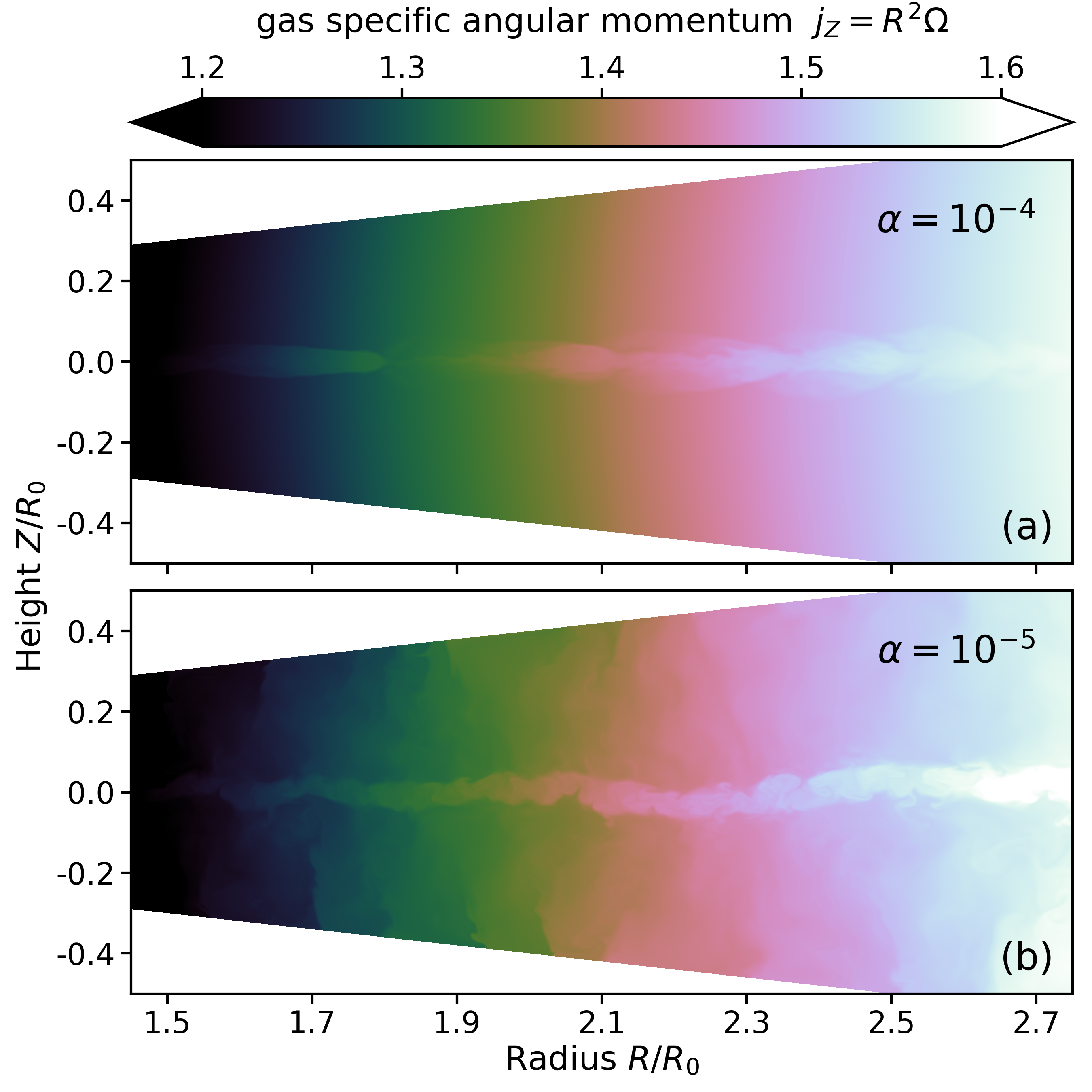}
\caption{Specific angular momentum of gas at $t = 400\,T_0$ from models with $\alpha = 10^{-4}$ (panel~$a$) and $\alpha = 10^{-5}$ (panel~$b$), both with ${\rm St}_0 = 10^{-3}$. The radial extent is the same as in Figs.~\ref{fig:stream_lines_1e-4a} and \ref{fig:stream_lines_1e-5a}.
\label{fig:compare_specificam}}
\end{figure}

Similar to Fig.~\ref{fig:stream_lines_1e-4a}, Fig.~\ref{fig:stream_lines_1e-5a} presents the dust-to-gas density ratio, but for the model with $\alpha = 10^{-5}$. In contrast to the higher-viscosity case, the shuttlecock-shaped substructures are absent in this model. Instead, the lower viscosity allows clearer signatures of VSI to emerge, particularly its corrugation modes, as illustrated by the streamlines in panel~$b$. The vertical gas motions reach typical Mach numbers of around 0.1, comparable to those reported for VSI-driven flows in inviscid models (\citealt{schafer.U.2020.03, huang.P.2025.06a}). These vertical motions corrugate the dust layer about the midplane, leading to discrepancies between $H_{\rm d}^{\rm rms}$ and $H_{\rm d}^{\rm std}$ in Fig.~\ref{fig:scale_height_1e-3s}$b$. 

The presence of VSI is further supported by the distribution of gas specific angular momentum shown in Fig.~\ref{fig:compare_specificam}. In the $\alpha = 10^{-5}$ model, pronounced vertical bands are evident in the disk's atmosphere, consistent with VSI-driven angular momentum mixing (\citealt{melonfuksman.JD.2024.02}). In contrast, such features are absent in the $\alpha = 10^{-4}$ model, supporting our interpretation that the VSI is suppressed at higher viscosity.

Despite the lower viscosity, which provides more favorable conditions for SI, the maximum dust-to-gas density ratio typically remains below 3, although our adopted metallicity is above the threshold $\mathcal{Z}_{\rm crit}(\alpha, \tau_{\rm s}) = 0.048$. Nevertheless, compared with the $\alpha = 10^{-4}$ model, a larger fraction of the disk midplane remains dense in the $\alpha = 10^{-5}$ case. This suggests that even a midplane with $\rho_{\rm d}/\rho_{\rm g} \gtrsim 1$ in a low-viscosity environment is insufficient to suppress the corrugation modes of the VSI.

In terms of the dust-layer thickness measured by $H_{\rm d}^{\rm std}$, the $\alpha = 10^{-5}$ model yields $H_{\rm d}/H_{\rm g} \gtrsim 0.25$ (see Fig.~\ref{fig:scale_height_1e-3s}$b$), comparable to the $\alpha = 10^{-4}$ case. However, this value is significantly larger than the prediction from Eq.~\ref{eq:hd_alpha} for $\alpha = 10^{-5}$. 

\subsection[Dust distribution in models with different alpha]{Dust distribution in models with different $\alpha$} \label{sec:compare_alpha_glob}

\begin{figure}[t]
\centering
\includegraphics[width = \columnwidth]{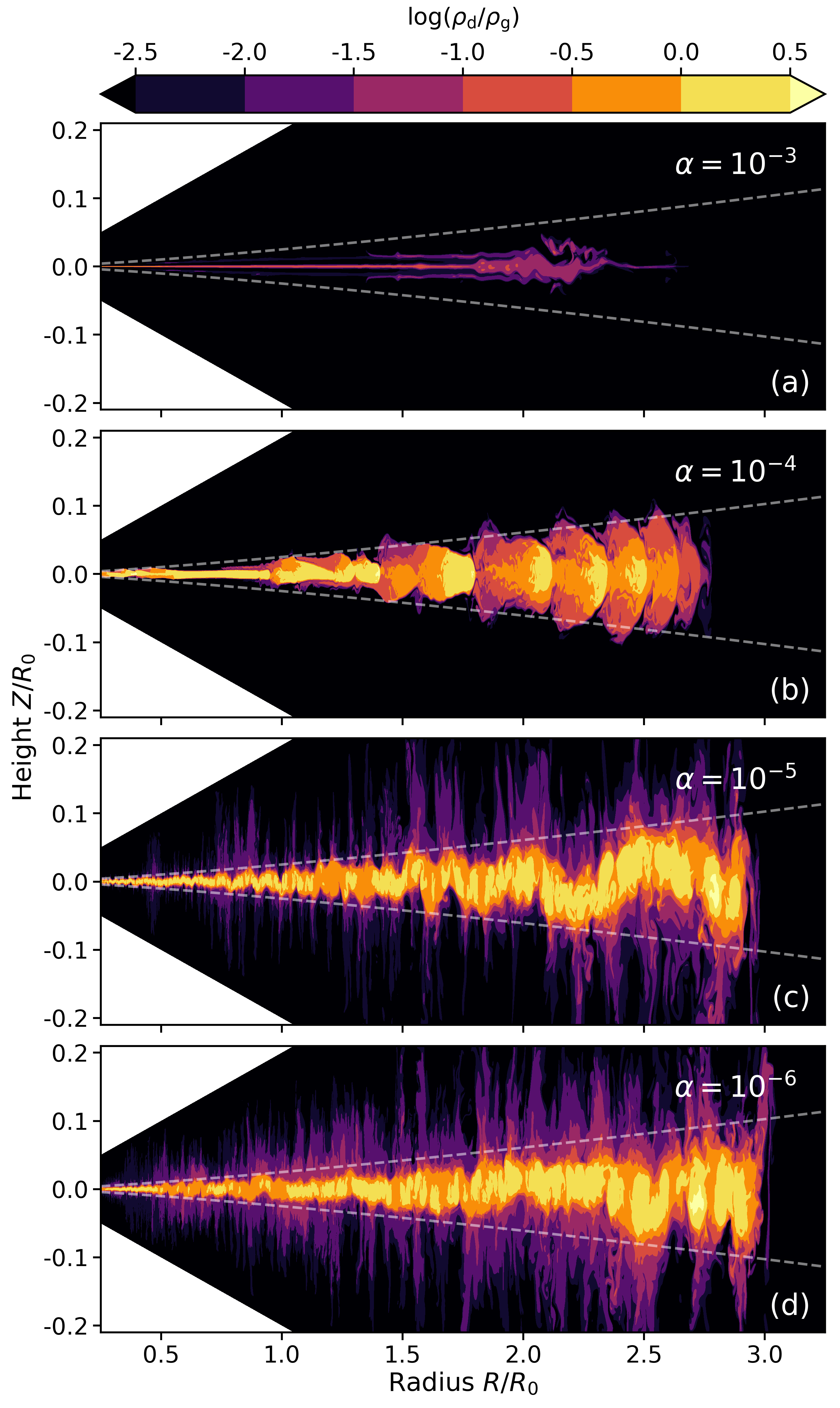}
\caption{Dust-to-gas density ratio at $t = 400\,T_0$ from models with different $\alpha$ values. Dashed curves denote contours for 25\% of the gas scale height.
\label{fig:compare_alpha_glob}}
\end{figure}

Given the distinct dynamical behaviors identified in Sects.~\ref{sec:stream_lines_1e-4a} and \ref{sec:stream_lines_1e-5a}, we now compare the dust distributions across these models, together with an additional run adopting an even lower viscosity of $\alpha = 10^{-6}$. The comparison is presented in Fig.~\ref{fig:compare_alpha_glob}.

The results indicate that shuttlecock-shaped substructures only arise in models where SI dominates the dust dynamics in an otherwise relatively undisturbed disk midplane. These structures disappear in higher-viscosity models where SI is suppressed, as well as in lower-viscosity models where active VSI interacts with SI in the midplane. 

For models with $\alpha \leq 10^{-5}$, the dust layer exhibits a qualitatively different morphology from those in higher-$\alpha$ models: the midplane becomes vertically corrugated due to VSI, while smaller-scale dust clumps produced by SI are still present. The primary distinction between panels~(c) and (d) is the increased prominence of vertical dust streaks in the lowest-viscosity model, reflecting the stronger VSI activity with lower $\alpha$.

\subsection{Dust distribution in models with different metallicities} \label{sec:compare_mass_ratio}

\begin{figure*}[t]
\centering
\includegraphics[width = \textwidth]{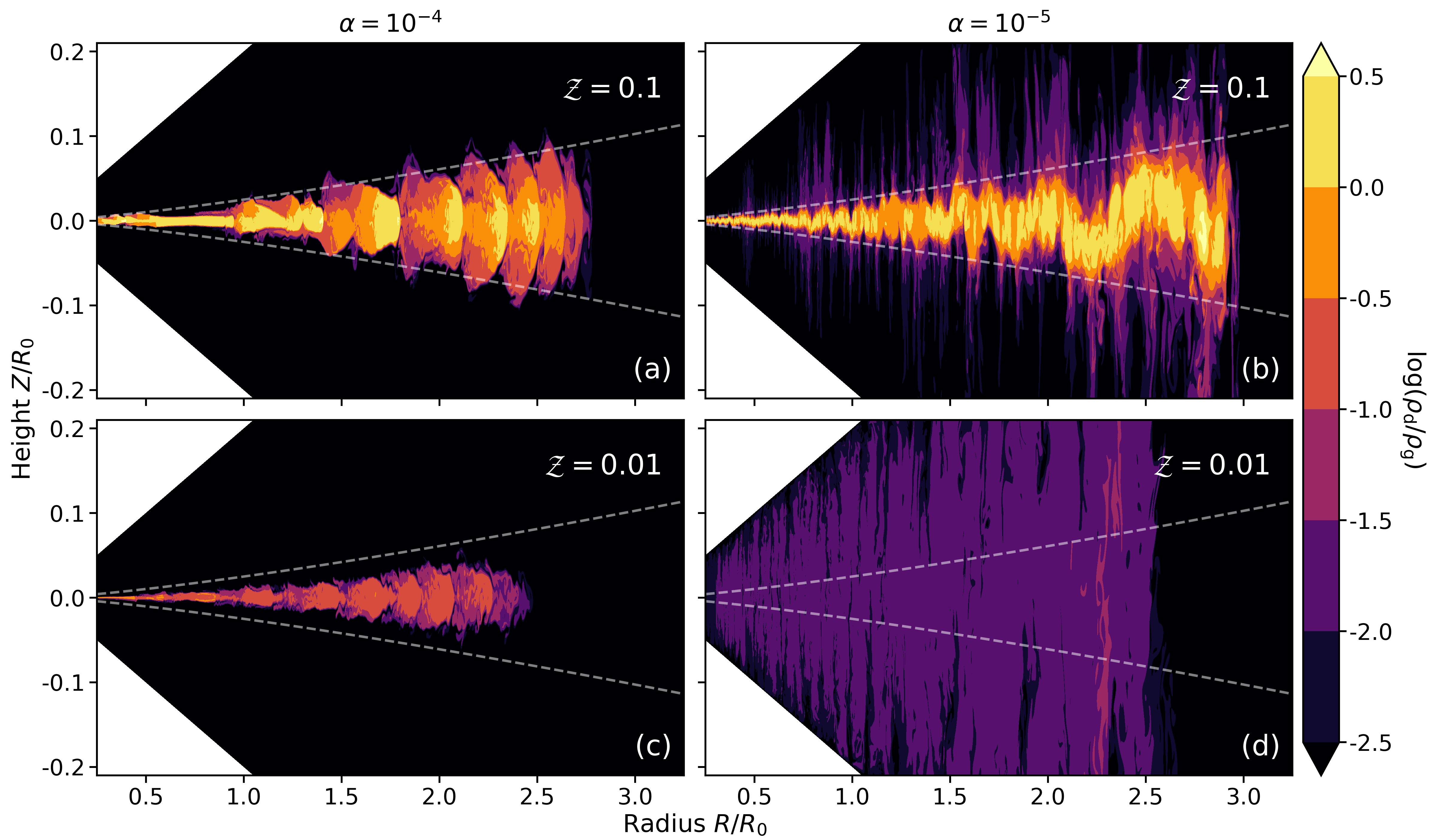}
\caption{Dust-to-gas density ratio at $t = 400\,T_0$ from models with different dust loads. The left and right columns correspond to models with $\alpha = 10^{-4}$ and $10^{-5}$, respectively. Upper panels: Results from the fiducial metallicity $\mathcal{Z} = 0.1$. Lower panels: Models with $\mathcal{Z} = 0.01$. Dashed curves denote contours for 25\% of the gas scale height.
\label{fig:compare_mass_ratio}}
\end{figure*}

In our fiducial models, we adopted a relatively high metallicity of $\mathcal{Z} = 0.1$ to ensure strong dust back-reaction. However, such an elevated metallicity may not be typical of protoplanetary disks. It is therefore important to assess how dust dynamics and the resulting substructures depend on the dust load. To this end, we performed two additional simulations at a lower metallicity, $\mathcal{Z} = 0.01$. The results are shown in Fig.~\ref{fig:compare_mass_ratio}.

A comparison between Figs.~\ref{fig:compare_mass_ratio}$a$ and \ref{fig:compare_mass_ratio}$c$ shows that the shuttlecock-shaped substructures persist even when the dust load is reduced to $\mathcal{Z} = 0.01$. However, the maximum dust-to-gas density ratio barely reaches unity, reflecting the lower metallicity. In addition, the vertical thickness of the dust layer decreases due to weaker SI, resulting from reduced dust back-reaction.

Comparing panels~$b$ and $d$, we find that while SI dominates the midplane dust dynamics in the $\mathcal{Z} = 0.1$ case, the weaker back-reaction at $\mathcal{Z} = 0.01$ allows VSI to dominate instead. As a result, the settled dust layer at the midplane is strongly disrupted and effectively destroyed.

\subsection{Dust distribution in models with different Stokes numbers} \label{sec:compare_grain_size}

\begin{figure*}[t]
\centering
\includegraphics[width = \textwidth]{figs/fig_compare_grain_size.png}
\caption{Dust-to-gas density ratio at $t = 400\,T_0$ from models with different reference Stokes numbers. The left and right columns correspond to models with $\alpha = 10^{-4}$ and $10^{-5}$, respectively. Upper panels: Results from the fiducial reference Stokes number ${\rm St}_0 = 10^{-3}$. Lower panels: Models with ${\rm St}_0 = 10^{-2}$. Dashed curves denote contours for 25\% of the gas scale height, and arrows below denote the Stokes number of dust grains in the midplane. The spatial domains differ between the upper and lower panels due to the different radial drift timescales.
\label{fig:compare_grain_size}}
\end{figure*}

\begin{figure}[t]
\centering
\includegraphics[width = \columnwidth]{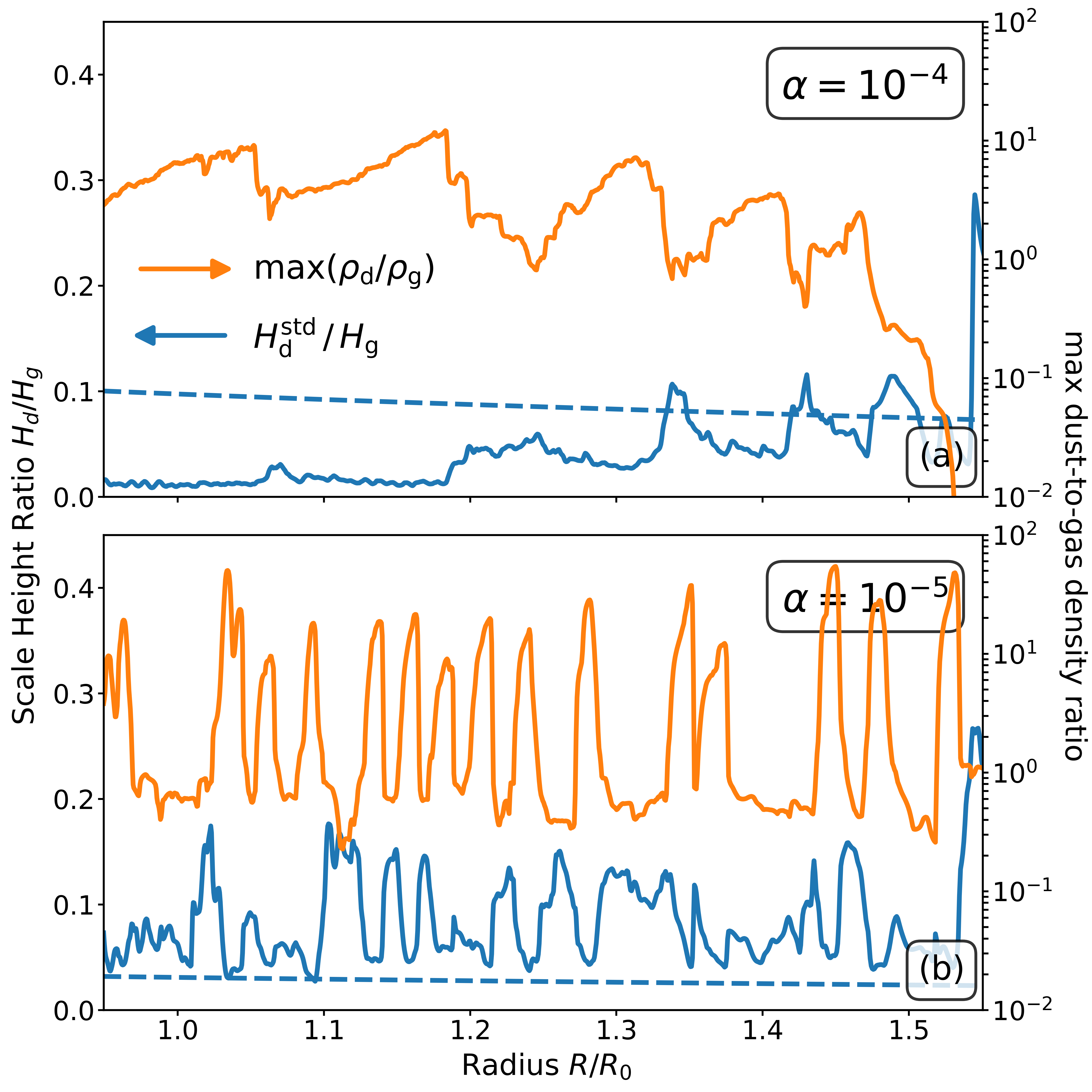}
\caption{Radial profiles of the dust-to-gas scale height ratio (left $y$-axis, blue curves) and the maximum dust-to-gas density ratio (right $y$-axis, orange curves) at $t = 400\,T_0$ from models with $\alpha = 10^{-4}$ (panel~$a$) and $\alpha = 10^{-5}$ (panel~$b$), both with ${\rm St}_0 = 10^{-2}$. The dust scale heights are measured using the standard deviation ($H_{\rm d}^{\rm std}$, Eq.~\ref{eq:hd_std}), and the dashed curve indicates a reference-only expected ratio assuming explicit dust diffusion, as given by Eq.~\ref{eq:hd_alpha}. The radial extent is the same as in Figs.~\ref{fig:stream_lines_1e-4a} and \ref{fig:stream_lines_1e-5a}.
\label{fig:scale_height_1e-2s}}
\end{figure}

So far, we have explored dust dynamics by varying $\alpha$ and $\mathcal{Z}$, the two key parameters entering the SI-clumping threshold $\mathcal{Z}_{\rm crit}(\alpha, \tau_{\rm s})$. However, we have not yet examined how dust dynamics and the resulting substructures depend on the grains' Stokes number (i.e., stopping time). To this end, we performed two additional simulations with ${\rm St}_0 = 10^{-2}$. The results are shown in Fig.~\ref{fig:compare_grain_size}, while measurements of the dust scale height and the maximum dust-to-gas density ratio are presented in Fig.~\ref{fig:scale_height_1e-2s}.

A comparison between Figs.~\ref{fig:compare_grain_size}$a$ and \ref{fig:compare_grain_size}$c$ shows that the shuttlecock-shaped substructures persist when the Stokes number increases to ${\rm St}_0 = 10^{-2}$. However, their vertical extent is significantly reduced due to the more efficient settling of larger grains. Nevertheless, the tail of the shuttlecock still exhibits a dust scale height larger than the value predicted by Eq.~\ref{eq:hd_alpha} (see Fig.~\ref{fig:scale_height_1e-2s}$a$). With the adopted metallicity $\mathcal{Z} = 0.1$ exceeding the threshold $\mathcal{Z}_{\rm crit}(\alpha, \tau_{\rm s}) = 0.054$ for ${\rm St} = 1.4\times10^{-2}$ at $R = 1.3R_0$, and given the enhanced settling of larger grains, the maximum dust-to-gas density ratio increases to typically $3 \lesssim {\rm max}(\rho_{\rm d}/\rho_{\rm g}) \lesssim 10$. However, this still falls short of the strong clumping regime.

Comparing Figs.~\ref{fig:compare_grain_size}$b$ and \ref{fig:compare_grain_size}$d$, the clumping efficiency becomes substantially higher, and the dust concentrations are more pronounced. Owing to the intrinsic turbulence generated by active SI in the midplane, the dust scale height in the $\alpha = 10^{-5}$ model is globally larger than the value predicted by Eq.~\ref{eq:hd_alpha} (see Fig.~\ref{fig:scale_height_1e-2s}$b$). In contrast to the smaller-grain cases, the midplane dust layer does not exhibit a prominent VSI-driven corrugated morphology, as larger grains are less tightly coupled to the gas. Furthermore, since the adopted metallicity $\mathcal{Z} = 0.1$ is well above the corresponding threshold $\mathcal{Z}_{\rm crit}(\alpha, \tau_{\rm s}) = 0.026$ for ${\rm St} = 1.4\times10^{-2}$ at $R = 1.3R_0$, SI operates in a strongly nonlinear regime. The maximum dust-to-gas density ratio reaches typical values of $20 \lesssim {\rm max}(\rho_{\rm d}/\rho_{\rm g}) \lesssim 50$, potentially exceeding the Hill density of the disk \citep{klahr.H.2020.09}:
\begin{equation} \label{eq:rho_hill}
\rho_{\rm Hill} = \frac{9}{4\pi}\frac{M_\star}{R^3}.
\end{equation}
For dust clumps at 100 au orbiting a solar-mass star, $\rho_{\rm Hill} = 4.3\times10^{-13}\,{\rm g/cm^3}$. Given a typical midplane gas density of $\rho_{\rm g} = 1.6\times10^{-14}\,{\rm g/cm^3}$ (corresponding to $\Sigma_{\rm g} = 6.0\,{\rm g/cm^2}$ and $H_{\rm g} = 10$ au), this implies that a dust-to-gas density ratio $\rho_{\rm d}/\rho_{\rm g} \gtrsim 27$ is required for gravitationally bound dust clumps. Our measured peak values therefore approach, and may exceed, this threshold. Meanwhile, we note this result does not contradict the conclusion of \citet{ostertag.D.2025.03}, who found that SI struggles to produce clumps exceeding the Hill density. The difference mainly reflects the orbital radii considered: the clumps in that work were studied near 10 au, whereas our clumps are located at 100 au. Since $\rho_{\rm Hill} \propto R^{-3}$ increases inward more steeply than the disk gas density, much stronger dust concentration is required at smaller radii. Using our disk model, a dust-to-gas density ratio of $\rho_{\rm d}/\rho_{\rm g} \gtrsim 139$ would be required at 10 au to reach the Hill density. The two results are therefore consistent and mainly reflect the strong radial scaling of the Hill density.

\section{Discussion} \label{sec:discussion}

\subsection{Limitations of the numerical setup}

The VSI arises in baroclinic disks where the angular velocity of gas varies with height owing to the radial temperature gradient \citep{nelson.RP.2013.11}. Studies have shown that the fastest-growing VSI modes are typically localized in the disk's surface layers, where the vertical shear is strongest, before extending toward the midplane \citep{stoll.MHR.2014.12,barker.AJ.2015.06}. Because of this intrinsically vertical character, previous numerical studies commonly adopt vertically extended domains, often spanning several gas scale heights (e.g., $\pm$4--5$H_{\rm g}$), to capture both the surface and body modes. The treatment of vertical boundary conditions is also known to influence VSI development, as wave reflection and transmission at the disk surfaces can modify growth rates and the nonlinear saturation amplitude \citep{wu.Y.2024.09}.

In our simulations, we adopted a vertically limited computational domain together with reflective vertical boundary conditions chosen for computational efficiency and numerical stability. We acknowledge that this restricted vertical extent may limit the VSI's full development, particularly for the fastest-growing surface modes. As a result, the saturated VSI amplitude, and hence the level of VSI-driven vertical stirring, may be underestimated in our models. This limitation should be borne in mind when interpreting the competition between VSI and SI and its influence on dust morphology. Future work employing more vertically extended domains and carefully designed boundary conditions will be necessary to fully capture the global VSI dynamics and its coupling to dust evolution.

We would also want to discuss the potential role of numerical diffusion in shaping the shuttlecock-shaped substructures identified in Sect.~\ref{sec:stream_lines_1e-4a}. Through private communications (e.g., Ostertag et al., in prep), we learned that SI in inviscid disk simulations evolved with relatively diffusive Riemann solvers can produce qualitatively similar dust morphologies, raising the question of whether the shuttlecock features might be numerical artifacts. We argue, however, that numerical diffusion is unlikely to dominate the behavior in our models. First, the vertical structure of the shuttlecocks, including both the head and the tail, exhibits a dust scale height larger than the reference value predicted by Eq.~\ref{eq:hd_alpha} (see Fig.~\ref{fig:scale_height_1e-3s}) and is resolved by $\gtrsim 100$ numerical cells. In the radial direction, the separation between neighboring heads also exceeds the dust scale height. Such spatial scales are therefore not expected to be strongly affected by numerical diffusion. Moreover, as shown in Appendix~\ref{sec:compare_integrator}, our cross-code comparison using the \textsc{pluto} code with both the Roe Riemann solver \citep{roe.PL.1981.10,mignone.A.2007.05}, one of the least diffusive solvers commonly used \citep{toro.EF.2009}, and the more diffusive Harten-Lax-van-Leer (HLL) solver yields results consistent with those obtained with \textsc{fargo3d}. This agreement, independent of the choice of Riemann solver, further supports the robustness of the shuttlecock structures.

We nevertheless acknowledge another limitation of our numerical setup: although gas viscosity is included, an explicit dust diffusion term is not implemented. This is because the public versions of all \textsc{fargo3d}, \textsc{pluto}, and \textsc{athena\plus\plus} codes currently lack a dust module that can simultaneously incorporate dust diffusion and consistently account for dust back-reaction onto the gas. Future studies would benefit from numerical frameworks capable of treating both processes self-consistently, such as the recently developed but not yet open-sourced dust fluid module for \textsc{athena\plus\plus} \citep{huang.P.2022.09}, which will enable more rigorous investigations of diffusion effects on coupled gas--dust dynamics.

\subsection{Outlooks}

Future work should adopt a more vertically extended domain to better capture VSI development and more rigorously assess its interaction with SI in viscous environments. Extending the models to three dimensions will also be essential to explore non-axisymmetric structures, such as zonal flows, vortices, and possible secondary instabilities arising from SI--VSI interactions. In addition, implementing numerical methods that consistently account for both dust diffusion and dust back-reaction will enable a more controlled investigation of the effects of diffusion on dust substructures.

We will also assess the feasibility of observing structures generated by VSI and SI and quantify their expected observational signatures. The potential observations will rely on dust grain alignment and the resulting net linear polarization of the thermal re-emission radiation, explicitly accounting for competing processes, in particular self-scattering \citep{lietzow-sinjen.M.2025.11}. Finally, we will derive quantitative observational requirements and compare them with the specifications of current and next-generation submillimeter and millimeter interferometric facilities.

\section{Conclusion} \label{sec:conclusion}

In this work, we investigated the radial and vertical distribution of dust in protoplanetary disks by performing two-dimensional, axisymmetric gas--dust simulations that include dust back-reaction and parameterized viscosity. Motivated by ALMA observations revealing ubiquitous dust substructures, we focused on understanding how the interplay between SI, VSI, vertical settling, and radial drift shapes the morphology and concentration of marginally coupled dust in disks. The results of our study can be summarized as follows:
\begin{itemize}[topsep = 6pt, itemsep = 6pt]

    \item In models with moderate viscosity $\alpha = 10^{-4}$, we identify a characteristic shuttlecock-shaped dust substructure composed of a dense, vertically settled ``head'' and a vertically extended, diffuse ``tail'' (Fig.~\ref{fig:stream_lines_1e-4a}). This morphology originates from the SI (Fig.~\ref{fig:compare_alpha_vert}), triggered by the back-reaction of marginally coupled dust concentrated near the midplane.

    \item For dust grains with $5 \times 10^{-3} \lesssim {\rm St} \lesssim 10^{-2}$, the dust-to-gas scale height ratio reaches $\gtrsim 0.25$ in the shuttlecock tails (Fig.~\ref{fig:scale_height_1e-3s}$a$), while the dust-to-gas density ratio attains $\lesssim 3$ in the heads (Fig.~\ref{fig:stream_lines_1e-4a}). For larger dust grains with $1.5 \times 10^{-2} \lesssim {\rm St} \lesssim 2.5 \times 10^{-2}$, the scale height ratio reaches $\gtrsim 0.1$ and the density ratio increases up to $\lesssim 10$ (Fig.~\ref{fig:scale_height_1e-2s}$a$). In both regimes, the maximum dust scale height exceeds the reference value predicted by the balance between turbulent diffusion and vertical settling (Eq.~\ref{eq:hd_alpha}), indicating strong vertical stirring induced by the SI. However, the maximum dust densities remain below the critical Hill density required for planetesimal formation (Eq.~\ref{eq:rho_hill}).

    \item In lower-viscosity models ($\alpha = 10^{-5}$), the shuttlecock morphology disappears, as the VSI becomes sufficiently vigorous to disrupt dust structures near the midplane (Fig.~\ref{fig:stream_lines_1e-5a}). We find that even a midplane dust-to-gas density ratio of order unity is insufficient to suppress VSI when a small but finite background viscosity is present. 

    \item The enhanced instabilities in lower-viscosity models lead to dust scale heights much larger than those predicted by Eq.~\ref{eq:hd_alpha}. For grains with ${\rm St} \gtrsim 10^{-2}$, the dust-to-gas density ratio can reach up to 20--50 (Fig.~\ref{fig:scale_height_1e-2s}$b$), potentially exceeding the Hill density and thus approaching conditions favorable for gravitationally bound clump formation.

\end{itemize}

\begin{acknowledgements} We thank the anonymous reviewer for the helpful comments. We also thank Pinghui Huang for insightful discussions on disk instabilities and dust dynamics. This work was supported by the Deutsche Forschungsgemeinschaft (DFG) under Grant No. 544937803. J.B. further gratefully acknowledges funding from COST Action CA22133 PLANETS, supported by COST (European Cooperation in Science and Technology). The simulations were carried out on computing clusters hosted by the Max Planck Computing and Data Facility (MPCDF). The data underlying this article will be made available upon reasonable request to the corresponding author.
\end{acknowledgements}

\bibliography{refs}
\bibliographystyle{aa}

\begin{appendix}

\onecolumn
\section{Numerical resolution test} \label{sec:compare_resolution}

\begin{figure*}[t]
\centering
\includegraphics[width = \textwidth]{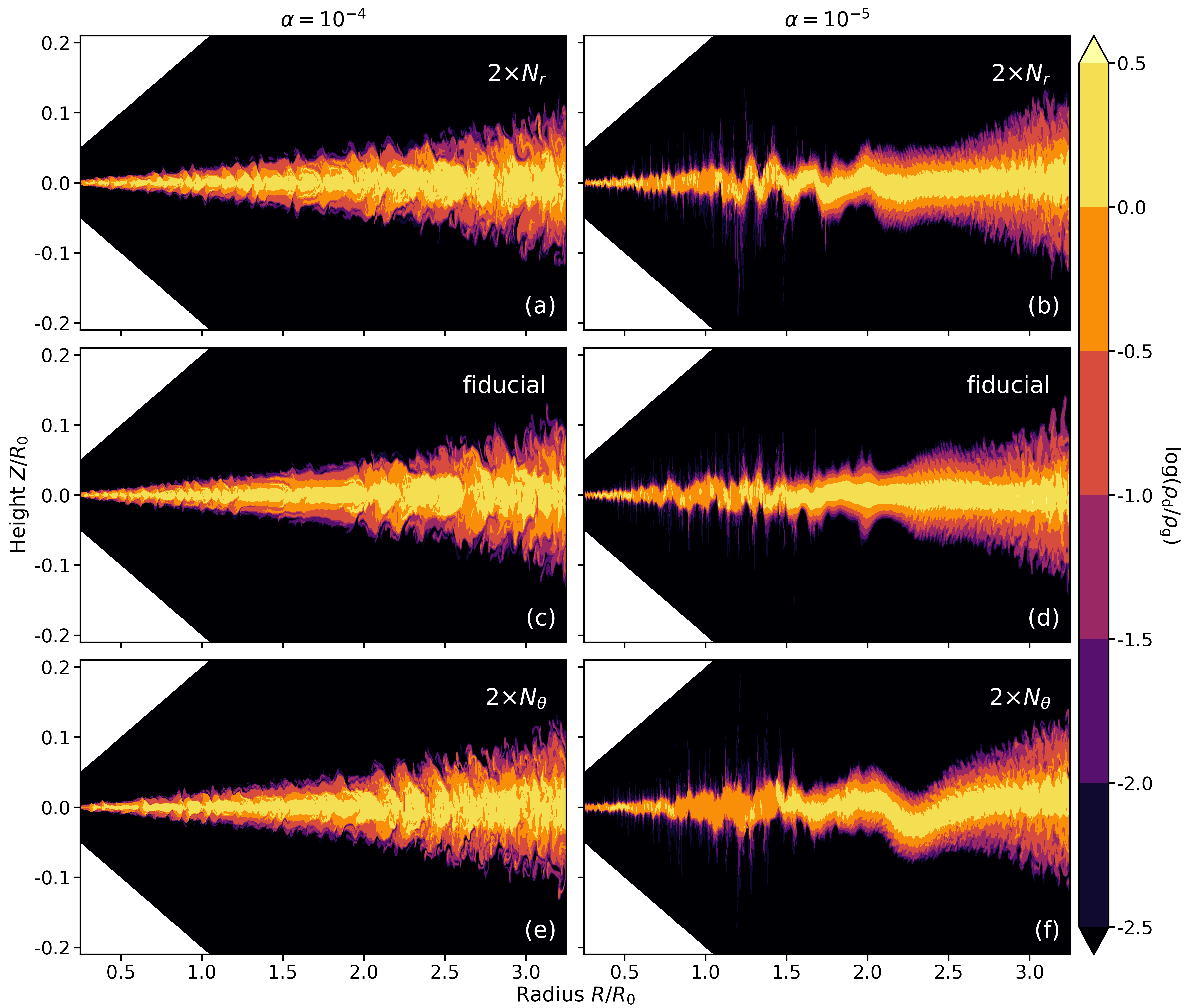}
\caption{Dust-to-gas density ratio at $t = 100\,T_0$ from models with different numerical resolutions. The left and right columns correspond to models with $\alpha = 10^{-4}$ and $10^{-5}$, respectively. Middle row: Models with the fiducial resolution (i.e., $\sim$200 cells per $H_{\rm g}$ in both directions). Top and bottom rows: Models with two times higher resolution than fiducial in the $r$ direction (top) and $\theta$ direction (bottom).
\label{fig:compare_resolution}}
\end{figure*}

To assess the robustness of our results against numerical resolution, we conducted a series of resolution tests in which the grid resolution was doubled separately in the radial and vertical directions for both the $\alpha = 10^{-4}$ and $\alpha = 10^{-5}$ models. The results are shown in Fig.~\ref{fig:compare_resolution}. Owing to the computational cost of higher-resolution simulations, these models were evolved for a shorter duration, $t = 100\,T_0$, which is sufficient to capture the initial development of the dust substructures. For the $\alpha = 10^{-4}$ models, we find good agreement in the overall dust morphology across all resolutions. In the $\alpha = 10^{-5}$ models, the higher-resolution runs, particularly the one with doubled vertical resolution, exhibit more pronounced midplane corrugations due to better-resolved VSI activity. Nevertheless, the fiducial simulation reproduces the key features of the dust distribution. These tests indicate that our results are numerically converged at the fiducial resolution and are not artifacts of grid discretization in either the radial or vertical direction. 

\onecolumn
\section[VSI activity in models with different alpha]{VSI activity in models with different $\alpha$} \label{sec:compare_alpha_long}

\begin{figure*}[t]
\centering
\includegraphics[width = \columnwidth]{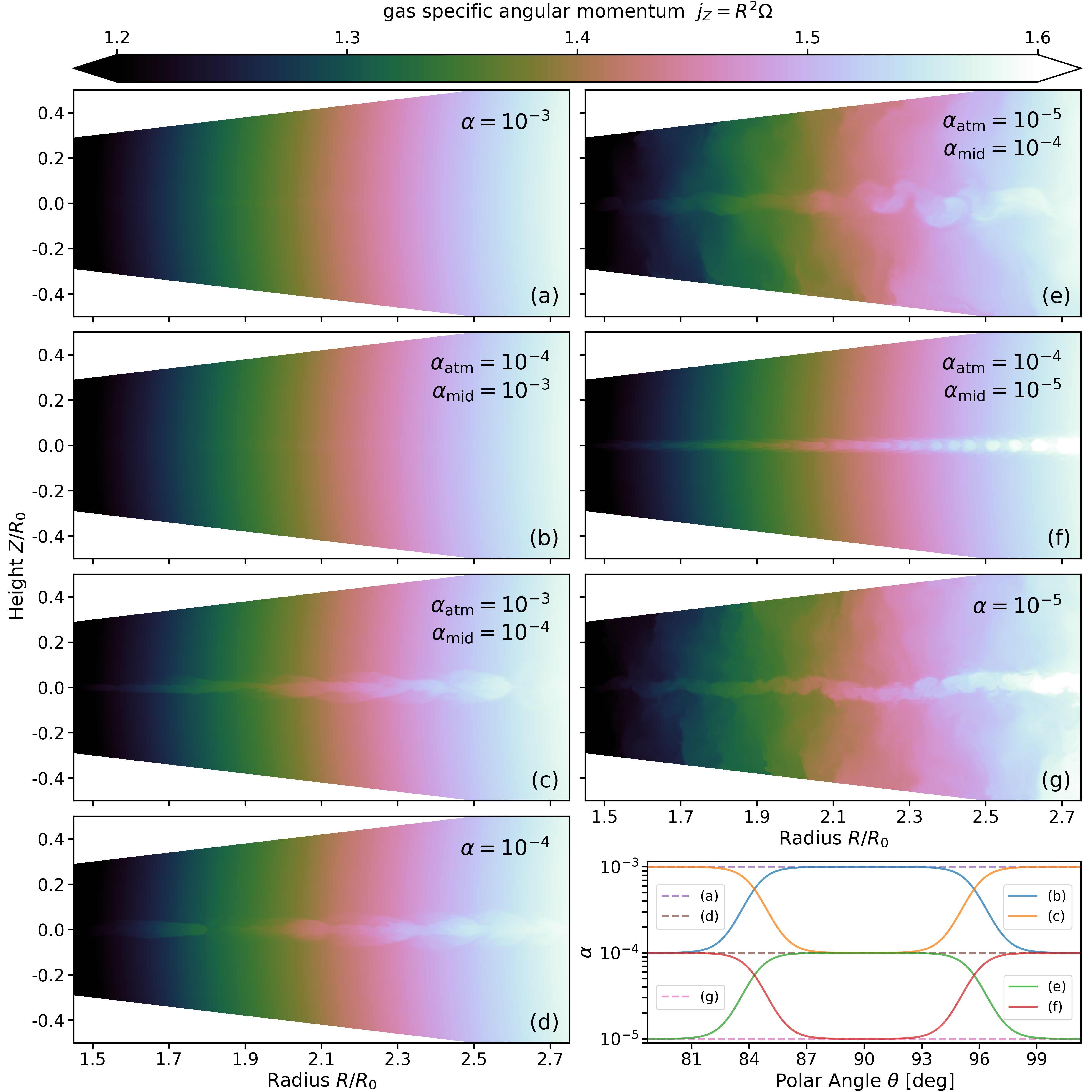}
\caption{Specific angular momentum of gas at $t = 400\,T_0$ from models with different vertical prescriptions of the $\alpha$ parameter. Panels~$a$, $d$, and $g$: Models in which $\alpha = 10^{-3}$, $10^{-4}$, and $10^{-5}$, respectively. Panel~$b$: Model in which $\alpha$ decreases from $10^{-3}$ in the midplane to $10^{-4}$ in the atmosphere. Panel~$c$: Model in which $\alpha$ increases from $10^{-4}$ in the midplane to $10^{-3}$ in the atmosphere. Panel~$e$: Model in which $\alpha$ decreases from $10^{-4}$ in the midplane to $10^{-5}$ in the atmosphere. Panel~$f$: Model in which $\alpha$ increases from $10^{-5}$ in the midplane to $10^{-4}$ in the atmosphere. Last panel: Corresponding vertical profiles of $\alpha$.
\label{fig:compare_spcam_long}}
\end{figure*}

\begin{figure*}[t]
\centering
\includegraphics[width = \columnwidth]{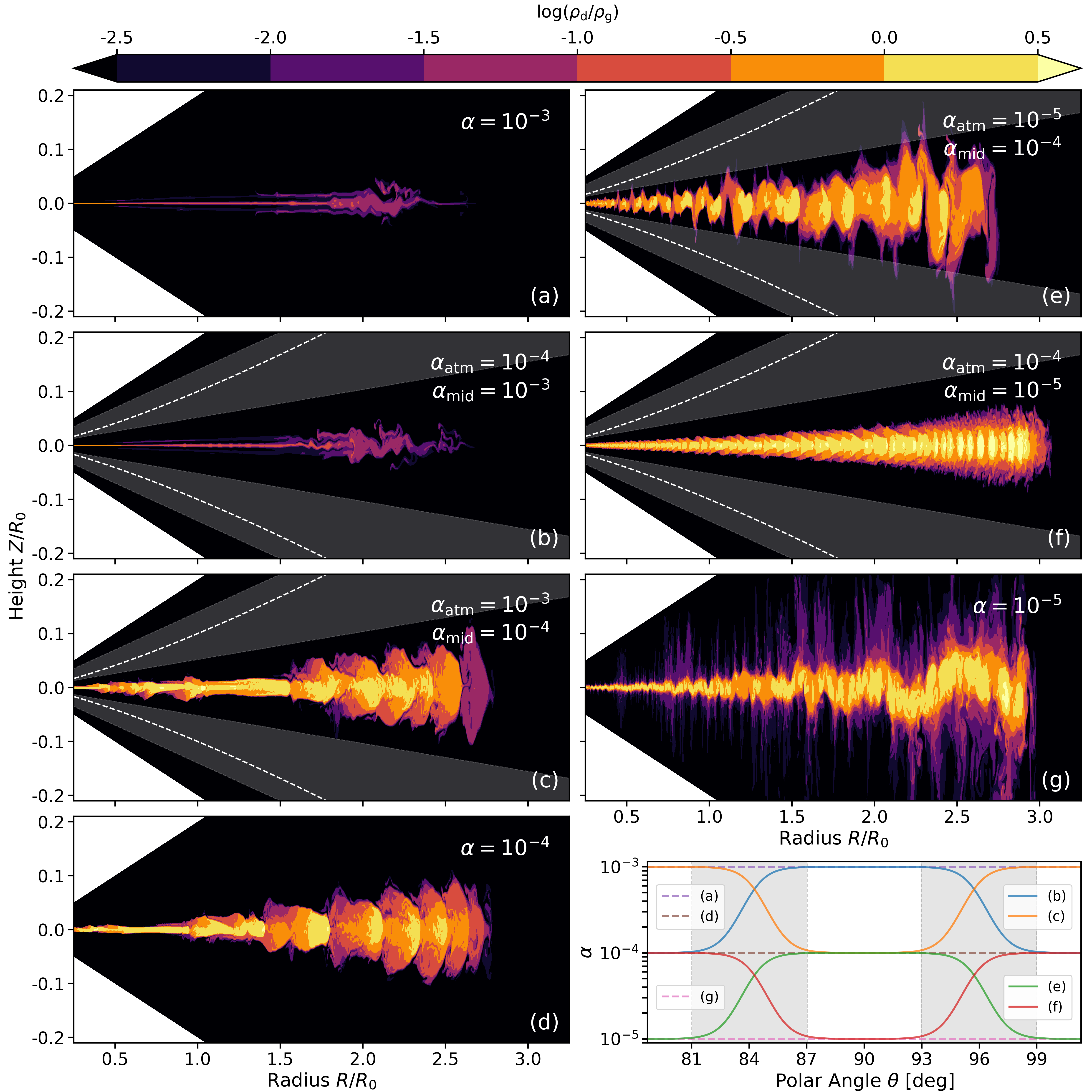}
\caption{Similar to Fig.~\ref{fig:compare_spcam_long} but for the dust-to-gas density ratio. Dashed curves in panels~$b$, $c$, $e$, and $f$ denote contours for one gas scale height. The shaded regions indicate where the transition of $\alpha$ occurs and are bounded by aspect ratios evaluated at the inner and outer radial boundaries.
\label{fig:compare_alpha_long}}
\end{figure*}

To examine the activity of VSI in our models and its role in shaping the dust layer, we performed two additional simulations in which $\alpha$ varies vertically between $10^{-4}$ and $10^{-5}$, similar to the setup explored in Sect.~\ref{sec:compare_alpha_vert}. The resulting gas specific angular momentum is shown in Fig.~\ref{fig:compare_spcam_long}, and the corresponding dust-to-gas density ratio is presented in Fig.~\ref{fig:compare_alpha_long}. We observe more pronounced signatures of VSI activity only in panels~$e$ and $g$ of Fig.~\ref{fig:compare_spcam_long}, which correspond to models with $\alpha = 10^{-5}$ in the disk atmosphere. This behavior is consistent with our earlier conclusion that the shuttlecock-shaped substructure is primarily driven by SI rather than VSI. Meanwhile, the comparison between panels~$d$ and $e$ of Fig.~\ref{fig:compare_alpha_long} shows that the midplane dust layer becomes more corrugated when $\alpha$ is reduced in the disk atmosphere, consistent with stronger VSI activity. The comparison between panels~$f$ and $g$ further shows that when $\alpha$ decreases in the disk midplane, the dust layer can form more clearly defined shuttlecock-shaped substructures (panel~$f$), although the morphology can also be significantly modulated by VSI-induced corrugations (panel~$g$).

\section{Dust substructures in VSI-suppressed models} \label{sec:compare_temperature}

\begin{figure}[t]
\centering
\includegraphics[width = 0.5\columnwidth]{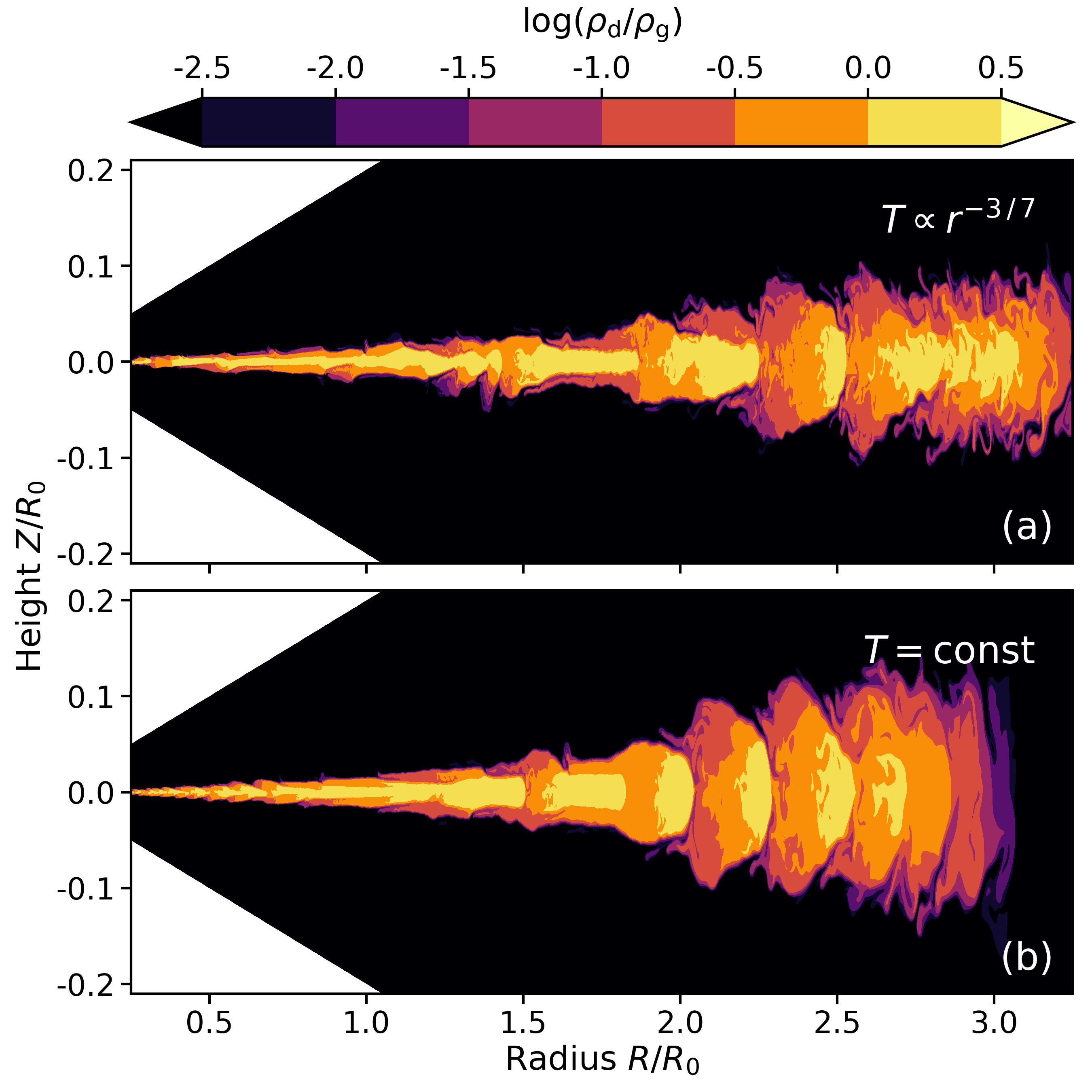}
\caption{Dust-to-gas density ratio at $t = 200\,T_0$ from models with our fiducial temperature profile $T\propto r^{-3/7}$ (panel~$a$) and with a radially isothermal prescription (panel~$b$). VSI is therefore completely suppressed in the latter case, and the resulting dust substructure is solely shaped by SI.
\label{fig:compare_temperature}}
\end{figure}

To further confirm that the shuttlecock-shaped dust substructures are driven primarily by the SI rather than the VSI, we performed an additional ``toy-model'' simulation at half the fiducial resolution in both dimensions, in which the temperature profile was modified to suppress VSI activity. Specifically, we imposed a radially constant temperature profile by setting the disk flaring index to 0.5. This removes the vertical shear in the gas rotation profile and thereby suppresses the VSI. The resulting dust-to-gas density ratio is shown in Fig.~\ref{fig:compare_temperature}. We find that the shuttlecock-shaped substructure still forms in this VSI-suppressed model, demonstrating that the SI alone can generate such dust morphologies in the absence of the VSI. This result further supports our conclusion that the dust substructures observed in our simulations are driven primarily by the SI rather than the VSI.

\section{Cross-code validation test} \label{sec:compare_integrator}

\begin{figure}[t]
\centering
\includegraphics[width = 0.5\columnwidth]{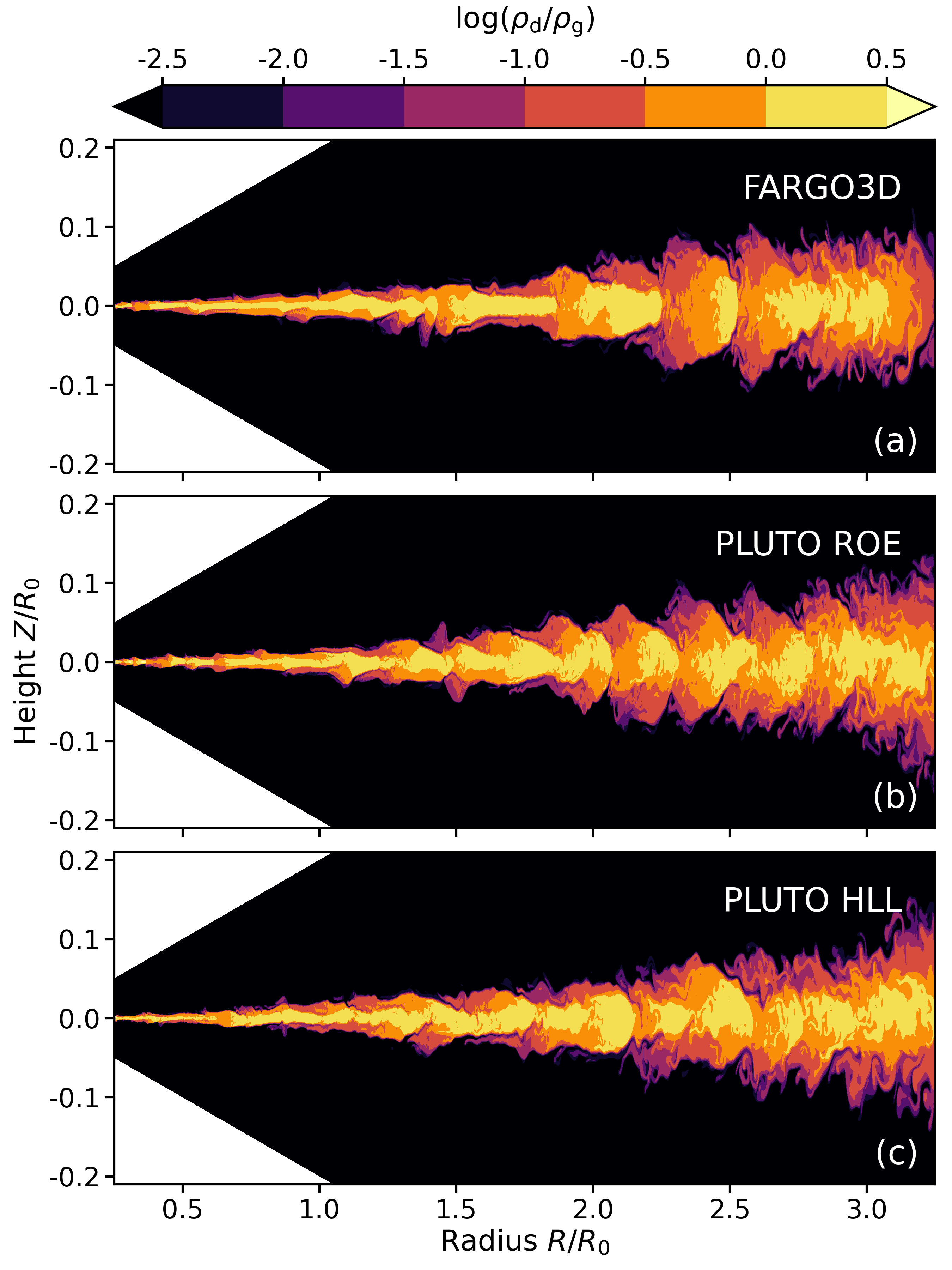}
\caption{Dust-to-gas density ratio at $t = 200\,T_0$ from models evolved with \textsc{fargo3d} (panel~$a$), the less diffusive Roe solver in \textsc{pluto} (panel~$b$), and the more diffusive HLL solver in \textsc{pluto} (panel~$c$).
\label{fig:compare_integrator}}
\end{figure}

To further verify the robustness of the results from the $\alpha = 10^{-4}$ model, we performed cross-code comparisons using the \textsc{pluto} code \citep{mignone.A.2007.05} and its dust-fluid module \citep{ziampras.A.2025.02} with the same physical setup described in Sect.~\ref{sec:methods}. The results of this comparison are shown in Fig.~\ref{fig:compare_integrator}.

To explore the potential impact of numerical diffusion, we carried out two \textsc{pluto} simulations with different numerical configurations. In the setup designed to minimize numerical diffusion, we adopted fourth-order parabolic reconstruction, third-order Runge-Kutta time integration for the gas, first-order implicit time integration for the dust, and the Roe Riemann solver \citep{roe.PL.1981.10,toro.EF.2009}. In the more diffusive configuration, we instead employed the HLL Riemann solver \citep{harten.A.1983.01}, replaced the parabolic reconstruction with linear, second-order reconstruction, and used second-order Runge-Kutta time integration for the gas.

Despite these differences, we find good agreement among the three models. In particular, the formation of the shuttlecock-shaped substructures and the magnitude of dust concentration are consistently reproduced. Minor quantitative differences appear in the detailed morphology and small-scale structures, likely due to differences in numerical methods. However, these variations do not affect the qualitative behavior or the main conclusions. This comparison demonstrates that our results are not specific to a particular numerical implementation and are robust across different simulation frameworks.

\end{appendix}
\end{document}